 \documentclass[preprint]{aastex}

\slugcomment{To appear in The Astrophysical Journal 000, 000Ð000 (2009)}

\shorttitle{Mid-IR variability in IC1396A}
\shortauthors{Morales-Calder\'on et al.}

\begin{document}


\title{Mid-Infrared Variability of protostars in IC 1396A}


\author{M.~Morales-Calder\'on\altaffilmark{1}, J.~R.~Stauffer\altaffilmark{2}, L.~Rebull\altaffilmark{2}, B.~A.~Whitney\altaffilmark{3}, D. Barrado y Navascu\'es\altaffilmark{1}, D.~R.~Ardila\altaffilmark{2}, I. Song\altaffilmark{4}, T. Y. Brooke\altaffilmark{4}, L.~Hartmann\altaffilmark{5}, and N.~Calvet\altaffilmark{5}}

\email{mariamc@laeff.inta.es}

\altaffiltext{1}{Laboratorio de Astrof\'{\i}sica Estelar y Exoplanetas (LAEX), Centro de Astrobiolog\'{\i}a (CAB, INTA-CSIC). LAEFF, P.O. 78, E-28691, Villanueva de la canada,  Madrid, Spain.}
\altaffiltext{2}{Spitzer Science Center, California Institute of Technology, Pasadena, CA 91125. }
\altaffiltext{3}{Space Science Institute, 4750 Walnut St. Suite 205, Boulder, CO 80301.}
\altaffiltext{4}{Astronomy Department, MC 105-24, California Institute of Technology, Pasadena, CA 91125.}
\altaffiltext{5}{Dept. of Astronomy, University of Michigan, 500 Church Street, Ann Arbor, MI 48109. }




\begin{abstract}
We have used Spitzer/IRAC to conduct a photometric monitoring program
of the IC1396A dark globule in order to study the mid-IR (3.6 - 8~$\mu$m)
variability of the heavily embedded Young Stellar Objects (YSOs) present in that area.   We
obtained light curves covering a 14 day timespan with a twice daily
cadence for 69 YSOs, and continuous light curves with approximately
12 second cadence over 7 hours for 38 YSOs.  Typical accuracies for
our relative photometry were 1-2\% for the long timespan data and
a few mmag, corresponding to less than 0.5\%, for the 7 hour continuous ``staring-mode" data.  More than half of
the YSOs showed detectable variability, with amplitudes from $\sim$0.05
mag to $\sim$0.2 mag. About thirty percent of the YSOs showed quasi-sinusoidal
light curve shapes with apparent periods from 5-12 days and light curve
amplitudes approximately independent of wavelength over the IRAC
bandpasses.   We have constructed models which simulate the time dependent
spectral energy distributions of Class I and II  YSOs in order to attempt to
explain these light curves.  Based on these models, the apparently
periodic light curves are best explained by YSO models where one or two high
latitude photospheric spots heat the inner wall of the
circumstellar disk, and where we view the disk at fairly large inclination
angle.  Disk inhomogeneities, such as increasing the height where the accretion funnel
flows to the stellar hotspot, enhances the light curve modulations. The other YSOs in our sample show a range of light curve
shapes, some of which are probably due to varying accretion rate
or disk shadowing events.  One star, IC1396A-47, shows a 3.5 hour
periodic light curve; this object may be a PMS Delta Scuti star.
\end{abstract}





\section{INTRODUCTION}

IC1396A (aka "The Elephant Trunk Nebula") is a prominent dark-globule
seen projected onto the bright nebular emission of the HII region
IC1396.  IC1396 is itself part of the larger Cepheus OB2 association,
which includes young clusters with ages ranging from ~10-12 Myr
(NGC 7160), to ~4 Myr (Tr 37), to $<$ 1 Myr (the proto-clusters forming
in globules such as IC1396A and IC1396N).   \citet{Reach04} reported
the first Spitzer observations of IC1396A, where they used IRAC to
identify three Class 0 or I protostars and a dozen Class II YSO's within
the boundaries of the globule.  \citet{Sicilia06} reported the results of a 
much wider area Spitzer IRAC and MIPS survey 
of the Cepheus OB2 association, where they derived disk frequencies
and Spectral Energy Distribution (SED) shapes for stars in Tr 37, NGC7160 and in IC1396A.  \citeauthor{Sicilia06} 
identified more than 50 YSO's in IC1396A, including 11 Class I stars
and 32 Class II YSO's.  They suspected that a number of other YSO's were
present in the field of view, but they could not be confirmed either because
there was too much nebular contamination of the photometry or because their 
observations did not allow them to identify Class~III sources (YSO's lacking IR excesses). 

Most of the IC1396A YSO population is heavily extincted, so there is
relatively little literature describing this proto-cluster.  Because
it is embedded in the IC1396 HII region, the normally assumed distance
to the globule is that adopted for IC1396 and Tr 37, which is 900 pc
\citep{Contreras02}.   \citeauthor{Sicilia06} compared optical photometry and
spectral type data for six of the most lightly reddened
globule members to \citet{Siess00} isochrones, and concluded that these
IC1396A stars have an average age of $\sim$1 Myr.  The more heavily
embedded members could presumably be even younger.

We have obtained time-series monitoring of the YSO's in IC1396A using
Spitzer's IRAC camera in order to (a) study the temporal variability of these
very young, heavily embedded stars and (b) identify additional members of
the globule population.  In Section 2, we describe the
observations and our photometric data analysis.  In Section 3, we 
provide new, deeper maps of the globule based on co-adding all of
our observations and from these images construct a new list of YSO
members of the globule population.  
Our time-series photometry and
identification of variable stars is described in Sec.~4, while Sec.~5 
includes a discussion of the different kinds of variability found and
possible physical mechanisms to explain the variability.
Finally, we summarize our findings in Sec.~6.

\section{OBSERVATIONS}

\begin{figure}[]
\plotone{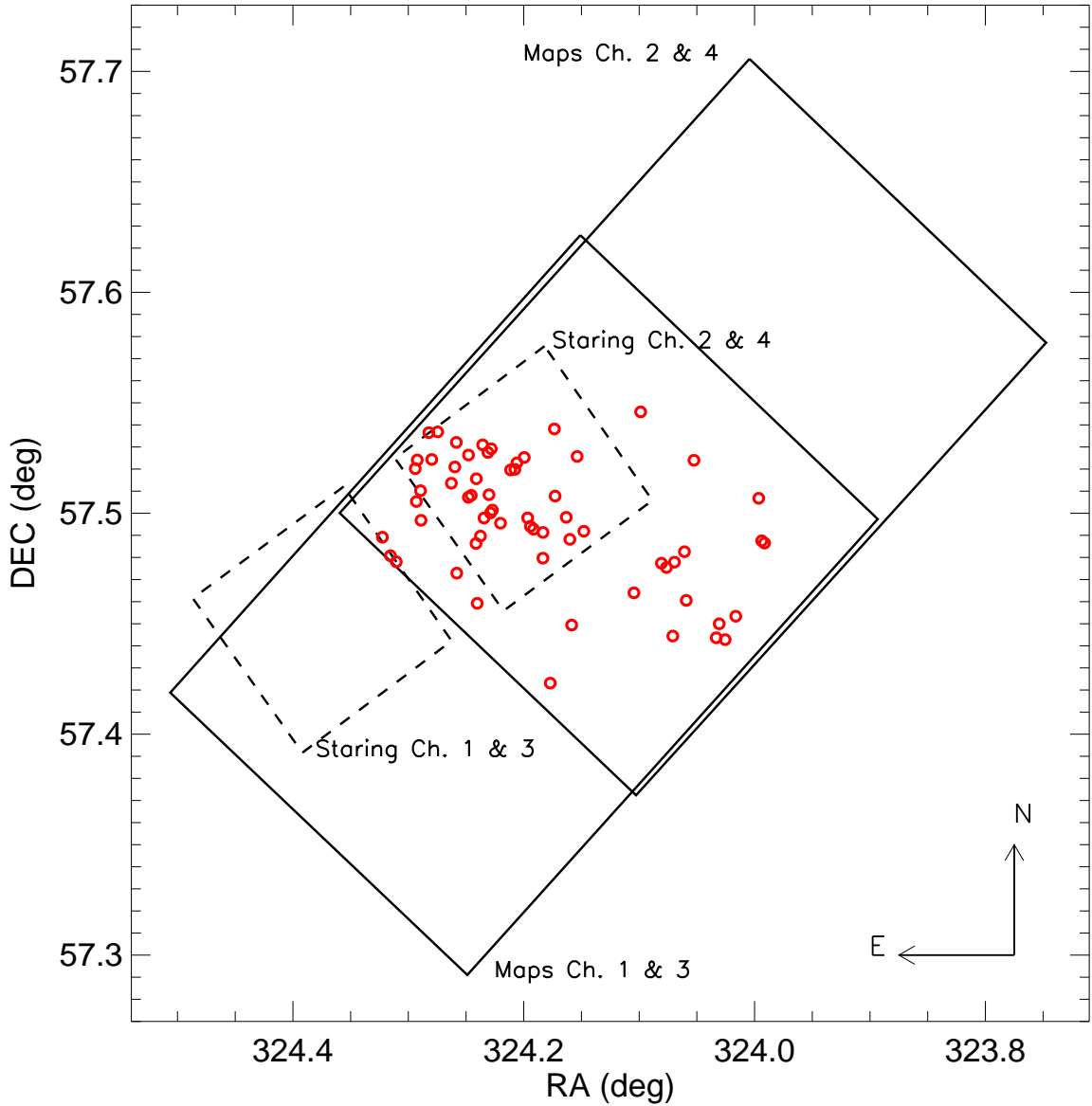}
\caption{Layout of observations. YSOs from \cite{Sicilia06} are marked with open circles. East is left, North is up. The solid and dashed rectangles show the field of view of the map and staring data respectively. All the previously known members are located within the area where mapping observations at Ch.~1 \& 3 and Ch.~2 \& 4 overlap.
}
\label{fig:layout}
\end{figure}

The goal of this program was to obtain well-sampled photometry for the 
protostars in IC1396A on timescales both of hours and days. 
Several previous programs have demonstrated that IRAC can provide extremely 
accurate and stable time-series photometry through the detection of the 
thermal emission of extra-solar planets \citep{Charbonneau05, Deming07} 
and the attempt to detect cloud formations in the photospheres of brown 
dwarfs \citep{Morales06}. Thus  we have used a recent DDT program (PI: 
Soifer, PID:470)  to study mid-IR variability of protostars in the very 
young star forming globule IC1396A. The data were collected from January 
24 to February 6, 2008 
and observations were performed with two distinctly different observing modes: 
\begin{itemize}
     \item{\bf Mapping mode:}
IRAC imaging was obtained for the whole IC1396A globule in mapping mode. 
About every 12~hours for 14 days, a short AOR was run to make a 2x3 map 
with individual exposures of 12 seconds ÒframetimeÓ (corresponding to 
10.4 second exposure
times) and a five position dithering at each map step. The IRAC maps do 
not cover the same FOV in all bands, providing a region of 
$\sim10\times11.5$', centered at  21:36:30.85 +57:29:49.37 with photometry 
in the four IRAC bands (see Figure~\ref{fig:layout}). 52 YSOs from 
\citet{Sicilia06} and 13 YSOs from \citet{Reach04} (11 of which are in 
common with the sample from \citeauthor{Sicilia06}) are included in this region.
     \item {\bf Staring mode:} In addition, on February 6,  a long staring 
AOR was performed. This AOR consisted of $\sim$7.5 hours of continuous monitoring 
with no dithering or mapping to get a
continuous set of photometry for a $\sim5\times5$' region. A total of 2000 
single observations, of 12 seconds frametime each, were performed. The Ch. 2 and 4 FOV was positioned at 
the core of the globule and it included 37 previously known YSOs.  The 
center of the Ch. 1 and Ch. 3 FOV is offset from the Ch. 2 and Ch. 4 FOV by of
order 7' to the SE. No previously known YSO's fall in that region. See 
Figure~\ref{fig:layout} for further information on the layout of observations. \\
\end{itemize}

We will use the staring data (Ch.~2 \& Ch.~4) to study the short term (from minutes to hours) 
variability and the mapping data (all channels) to analyze longer term (up to 14 days) variability. In addition, the mapping observations will be combined to produce new deeper maps and search for new members of the globule.\\

%
\subsection{Mapping Data Photometry} \label{mapping_observations}
%
%

\begin{figure}[]
\epsscale{.3}
\plotone{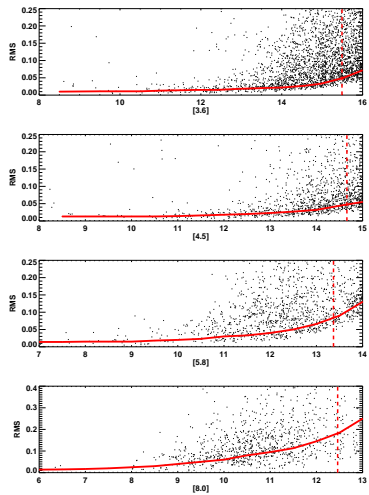}
\caption{Light curve RMS versus mean magnitude for all detections in the mosaic at 3.6~$\mu$m. The fitted polynomial (red solid line) has been used as a measure of the error. The vertical dashed line represents the magnitude of our faintest target.}
\label{fig:rms}
\end{figure}

Our starting point for the data analysis was the post-Basic Calibrated Data 
(postBCD) mosaic image produced by the IRAC pipeline software (ver. S17.0.4). 
All the data have been analyzed with IRAF standard procedures. 
We performed aperture photometry using PHOT with a source aperture of 3 
pixels radius (3.''60). The aperture radius was selected in order to obtain 
the maximum signal-to-noise ratio. The sky background was subtracted using 
a 4 pixels (4.''80) wide annulus. The corresponding aperture corrections, 
taken from the {\it Spitzer} web page  have been applied. The mean exposure 
time for the final mosaics is 52 seconds and we have a total of 28 exposures 
taken every 12 hours.
Calculating appropriate photometric errors is crucial in order to apply 
variability tests such as the $\chi^2$ test and thus we decided to compute 
the photometric errors empirically from the data themselves. 
Figure~\ref{fig:rms} plots the observed photometric rms in the time series 
for each star detected at 3.6, 4.5, 5.8, and 8.0~$\mu$m respectively, as a function of magnitude. We used 
a polynomial fit close to the lower envelope for each IRAC channel as 
the  estimated photometric uncertainty for a given magnitude.

\begin{figure}[]
\epsscale{.3}
\plotone{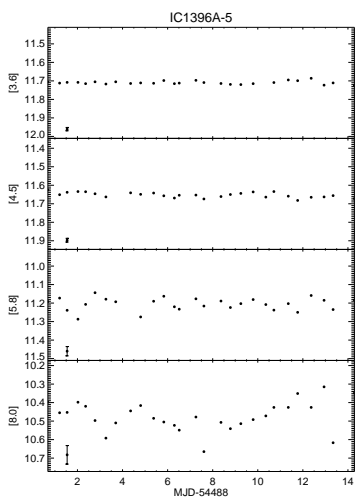}
\caption{Example of a light curve for a non-variable YSO, IC1396A-5, in the four IRAC bandpasses.
The RMS of the light curves is 9, 14, 34 and 80~mmag for IRAC channels 1 through 4 respectively.}
\label{fig:maps}
\end{figure}

An example of the time series for one of our non-variable targets 
can be seen in Figure~\ref{fig:maps}.

\subsection{Staring Data Photometry} \label{staring_observations}
%

In the case of the Staring data, our starting point was the individual 
Basic Calibrated Data (BCD) frames produced by the IRAC pipeline software 
(ver. S17.0.4) at the Spitzer Science Center (SSC). This pipeline produces 
fully flux-calibrated images, in units of MJysr$^{-1}$, that have had most 
of the instrumental signatures removed.
Some artifacts remain in the pipeline processed data, however - so we
have used IDL code provided by the Spitzer Science Center to correct
for column pull down and muxbleed (Ch.~1 and 2 features produced  by the presence of very bright sources). 

As for the mapping data, aperture photometry was performed in each image 
using PHOT with an aperture of 3 pixels radius, a background annulus of 
4 pixels width and the appropriate aperture corrections\footnote{The BCD
data have the native pixel scale of about 1.22"/pixel and have not been
distortion corrected, so the standard aperture correction is only 
approximately correct; however, since our interest is only in photometric
variability the small errors this introduces are irrelevant.}.
In order to improve the signal-to-noise ratio, the BCD images were combined in groups 
of 5. Therefore, we have 400 merged data points with 1 minute increments 
spanning almost 7.5~hr of observation time.

\begin{figure*}[]
\plotone{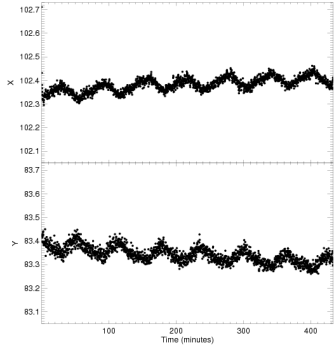}
\caption{Array x-position (top) and y-position (bottom) as a function of time for one of our targets. Only the already known 3000 sec oscillation of the pointing can be seen. }
\label{fig:xy}
\end{figure*}

Figure~\ref{fig:xy} illustrates the pointing stability during these observations. Only a very 
small (less than 0.3") periodic movement is seen. This effect is 
systematic and documented in the {\it Spitzer} web page. Such motion can affect the Ch. 2 fluxes at of order the 1\% level;
however, because we know the pointing movements precisely, we can
identify any photometric variations related to this effect.   Because we see
no measureable variations on the timescale of the pointing drift, we have
not attempted to correct our photometry for the ``pixel-phase" effects. 
The uncertainties in the light
curves were computed empirically. We assumed
that no significant real variability in our objects occurs on timescales
of 10 minutes or less. We measured the scatter of every
10 data points, and the 1$\sigma$ error bars in the figures
represent the median of these values.

The brightest objects of our sample show an upward trend
in brightness of $<$3\% from the beginning to the end of the observation
at Ch.~4. Because the same trend cannot be seen in the Ch.~2 data and because this variation only appears in the brightest objects, we believe that it is due to the latent image charge buildup, an instrumental 
effect that was also observed in previous cases with similar kinds of data 
\citep{Charbonneau05, Morales06}.
This instrumental effect may depend on the flux of the target and there 
is also a pixel-dependent term in the behavior of the long-term latents.
In 
addition, it is possible that they are frame time dependent. 
To derive a first-order correction for this effect, we have selected a
bright, non-variable star in our data and fitted a second-order polynomial
to its normalized flux.  We approximately correct the fluxes for all other bright stars (approx. [8.0]$<$10) which show this trend,
by dividing their time series photometry by this polynomial fit.

\section{COMBINED MAPS}
\begin{figure*}[]
\plotone{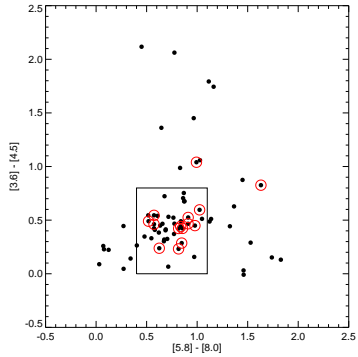}
\caption{IRAC color-color diagram after \citet{Allen04}. Black filled circles represent the whole sample of YSOs and red open circles stand for our new Class~I and Class~II selections. Objects inside the box are classified as Class~II objects while redder colors are indicative of Class~I objects and objects close to (0,0) are either field objects or Class~III sources \citep{Allen04}.
}
\label{fig:IRACCCD}
\end{figure*}

%
%
In addition to the individual mosaics, we have co-added all the mapping mode 
observations using MOPEX in order to produce a new deeper map.  The photometry 
has been obtained in the same way as 
in the previous cases and we have used the IRAC color-color diagram, 
[3.6]-[4.5] vs. [5.8]-[8.0], to target possible candidates and analyze 
their time series too. Thus, we are requiring our new candidate members to have detections at all IRAC bands. This color-color diagram was presented as a tool to separate 
young stars of different classes \citep{Allen04,Megeath04}, depending on 
where they fall in the diagram. Note however that Class~III objects cannot 
be separated from field objects due to their lack of infrared excess. The 
Elephant Trunk Nebula is embedded in a web of bright, spatially variable nebulosity which 
is brightest at 8~$\mu$m while normal stars are usually faint at that 
wavelength. Thus, we have excluded from our list of new candidates those 
objects showing large [5.8]-[8.0] but small [3.6]-[4.5] colors, and that are 
located within the brightest parts of the nebulosity or the parts with 
largest gradients between wavelengths.  In addition, we have eliminated from our list all the objects with errors larger than 0.1~mag in one or both of the bluer IRAC bands in order to have a reliable sample of new candidate members. Figure~\ref{fig:IRACCCD} shows the 
IRAC [3.6]-[4.5] vs. [5.8]-[8.0] diagram with the previously known YSOs 
as well as our new candidates plotted.  

\begin{figure*}[]
\epsscale{.5}
\plotone{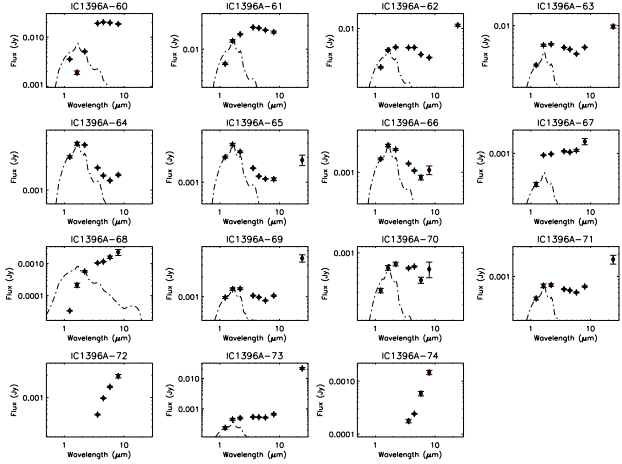}
\caption{Spectral energy distributions for our new candidates. All of them have been classified as Class~II or Class~I objects by means of the IRAC color-color diagram and have thick disks surrounding the central object. The J, H, and $K_s$ magnitudes come from the 2MASS database \citep{Cutri03}. The dashed lines are Kurucz models \citep{Castelli97} aimimng to represent the photosphere of each target assuming that the bluest 2MASS data point with good quality photometry represents the photospheric flux.
}
\label{fig:seds}
\end{figure*}

In addition, we have used the spectral energy distribution (SED) of each source as a way to characterize the structure of its
circumstellar disk. After \cite{Lada06} we have used the 3.6-8.0~$\mu$m slope for 
each source detected in all four IRAC bands to distinguish between 
objects with optically thick, primordial disks ($\alpha_{IRAC} >$ -1.8), objects surrounded by 
optically thin or anemic disks (-2.56 $< \alpha_{IRAC} <$ -1.8) and objects without disks 
($\alpha_{IRAC} <$ -2.56). This is based on the distribution of the disk
population in Taurus (1-2 Myr). The SEDs of our 
new targets can be seen in Figure~\ref{fig:seds}. We have used previous observations from the Spitzer archive to derive the fluxes of our new targets at 24~$\mu$m (PID:58, PI:Rieke). Most of our new candidates are located within the head of the globule, where the emission at 24~$\mu$m is very intense and thus we were able to derive magnitudes for only a few of our targets. Our new candidate members with their photometry and assigned evolutionary class 
(by means of the IRAC color-color diagram) are provided in
Table~\ref{tab:newphot}.  All these objects have thick 
disks based on their IRAC SED slope.

\section{TIME SERIES AND PRESENCE OF VARIABILITY}
%
%

We have tabulated and examined the time series photometry for all of
the objects detected in the mapping and staring data.
1$\sigma$~rms uncertainties of $\sim$2~mmag in Ch. 
2 and $\sim$4~mmag in Ch. 4 for the brightest targets are achieved for 
the staring data. For the mapping data the lowest uncertainties achieved 
are of order 6~mmag in the two bluest bandpasses. The light curves of all 
previously known and new candidate members 
were visually inspected. 
In this way different types of variability, 
that may be difficult to pick out in an automated fashion, can be spotted 
while bad photometric measurements can
be eliminated. However, this process is totally subjective. Since we wanted 
to search for variability also in the field, and given the large number 
of detections, we have used the $\chi^2$ test and the Stetson J statistic as quantitative 
estimators of photometric variability for both short- and long-term variability.

The $\chi^2$ test determines the probability that the deviations in a light curve
are consistent with the photometric errors (i.e., nonvariable). The
null hypothesis for the test is that there is no variability. We
evaluated the  $\chi^2$ statistic,
\begin{equation}
\chi^2=\sum_{k=1}^{k=K} \left(\frac{\Delta m(k)}{\sigma}\right)^2
\end{equation}
where $K$ is the number of data points in the light curve, $\Delta$m($k$) 
is the magnitude for each data point with the mean magnitude
subtracted, and $\sigma$ is the rms error in the phototmetry.
A large $\chi^2$ value indicates a greater deviation compared to
the photometric errors and thus a smaller probability that the null hypothesis
is true (i.e., variable). This probability, p, is calculated, and we
claim evidence for variability if  p $<$ 0.01.
Because this method is very sensitive to the accuracy of the error
estimation, our technique of using the data to empirically estimate the
errors should minimize false detections. In addition,  since a single bad measurement can produce 
high $\chi^2$ values, we eliminated from the time series the 3$\sigma$ 
isolated deviants prior to calculating the statistic.

\begin{figure*}[]
\epsscale{.5}
\plotone{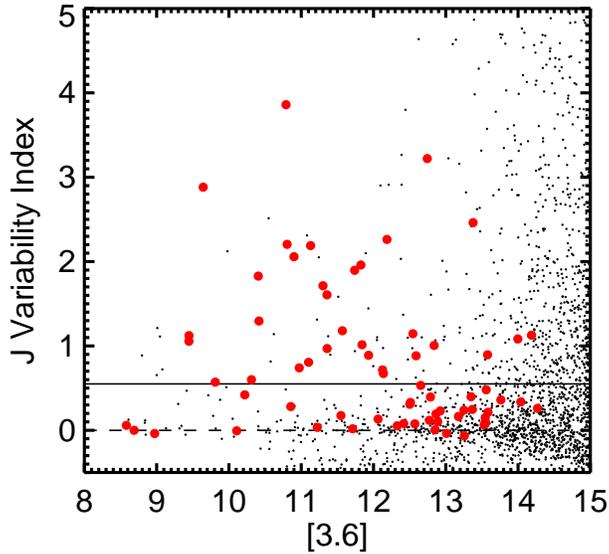}
\caption{Stetson variability index (J) plotted as a function of [3.6] mean magnitude
for stars brighter than [3.6]=15. The
dashed line at J=0 shows the expected value of the variability index for
non variable stars, and the dotted line at J=0.55 represents the minimum
adopted value used to identify variable stars in this study. Our IC1396A YSOs are represented with red filled circles.}
\label{fig:stetson}
\end{figure*}

The $\chi^2$ test does not take advantage of correlated changes in multiband time series and thus we used the Stetson variability index, J,  \citep{Stetson96} to try to identify low amplitude, correlated  variables. The Stetson variability index was computed for each star from the four IRAC bands magnitudes and  their associated photometric uncertainties as
\begin{equation}
J=\frac{\sum_{k=1}^{n}w_ksgn(P_k)\sqrt{\mid P_k \mid}}{\sum_{k=1}^{n}w_k}
\end{equation}
where we assign, to each of the n pairs of observations considered, a weight w$_k$, and 
\begin{equation}
P_k=\cases{\delta_{i(k)}\delta_{j(k)}~~~if~i(k) \ne j(k) \cr \delta_{i(k)}^2 - 1 ~~~~if~i(k) = j(k)}
\end{equation}
is the product of the normalized residuals of the two paired observations i and j, and
\begin{equation}
\delta_i=\sqrt{\frac{n}{n-1}}\frac{m_i-\overline{m}}{\sigma_i}
\end{equation}
is the magnitude residual of a given observation from the average normalized by the standard
error. 
For a non variable star, with only random noise,
the Stetson variability index should be scattered around
zero and have higher, positive values for stars with correlated
physical variability.
Figure \ref{fig:stetson} shows the Stetson statistic as a function of the
[3.6] magnitude. The dashed line at J = 0 shows the expected value of the variability index 
 for non variable stars, and the dotted line at J=0.55 represents the minimum value adopted by \citet{Carpenter01} in a  similar near-IR variability study toward a several square degree 
area in the Orion. Most of our targets have magnitudes brighter than [3.6]=13.5 and only one (catalogued by both the $\chi^2$ test and the J statistic as non variable) has a magnitude fainter than [3.6]=14.2. Following the choice
adopted by \citet{Carpenter01}, and based on our own comparison of the $\chi^2$ and J statistic results for individual stars, we adopt J = 0.55 as the borderline between variable and non variable objects. However, we note that the J index shows higher dispersion towards fainter magnitudes  and a different minimum value should be used for fainter objects.   

If evidence of variability was found in an object, we looked
for a periodic signal in the data following the methodology described
by \citet{Scargle82}. This method is equivalent to a least squares
fit (in the time domain) of sinusoids to the data. The
algorithm calculates the normalized Lomb periodogram for the
data and gives us a false-alarm probability based on the peak
height in the periodogram as a measure of significance. There is, in 
general a good agreement between the periods found at different wavelengths, 
however caution should be taken since in most cases our data only cover a time frame of order the
estimated period.

\section{RESULTS AND DISCUSSION} 
%
%

\subsection{ The mapping data: long-term variable members}\label{longtermvariability}
%
%

Fifty two previously known YSOs from \citet{Sicilia06} and 13 YSOs from \citet{Reach04} (11 of them are in common with the sample from \citet{Sicilia06}) plus 15 new candidates fall inside the area of 
our mosaics, and most of them have good photometry in the four IRAC bands.   As the first step
to determine their variability, we applied the $\chi^2$ test to the Ch.~1 \& Ch.~2
data.  We did not calculate $\chi^2$ for the longer wavelength channels because
those data are much noisier. Forty stars were labeled by the  $\chi^2$ 
statistics as variable objects in IRAC Ch.~1 and Ch.~2 while 4 objects, 
IC1396A-62, IC1396A-63, IC1396A-66, and IC1396A-72, were found to be variable only in Ch.~1. 
Two of these objects, IC1396A-62 and IC1396A-63 are located near the border of the 
Ch.~2 \& Ch.~4 mosaics and are outside the FOV in most of them; however, 
the light curve built for these objects with the photometry of Ch.~3 confirms 
the variation shown in Ch.~1. IC1396A-66 shows low amplitude 
variations in Ch.~1 but the light curves at longer wavelengths, even when they follow the same trend, are too 
noisy to confirm the variation. 
Object 72 has been detected in the deep mosaics which have been built combining all the individual maps, however in the individual observations only the Ch.~1 data is good enough to build its time series. We thus cannot confirm the Ch.~1 result, and consider
object~72 non-variable.

\begin{figure}[]
\epsscale{0.8}
\plotone{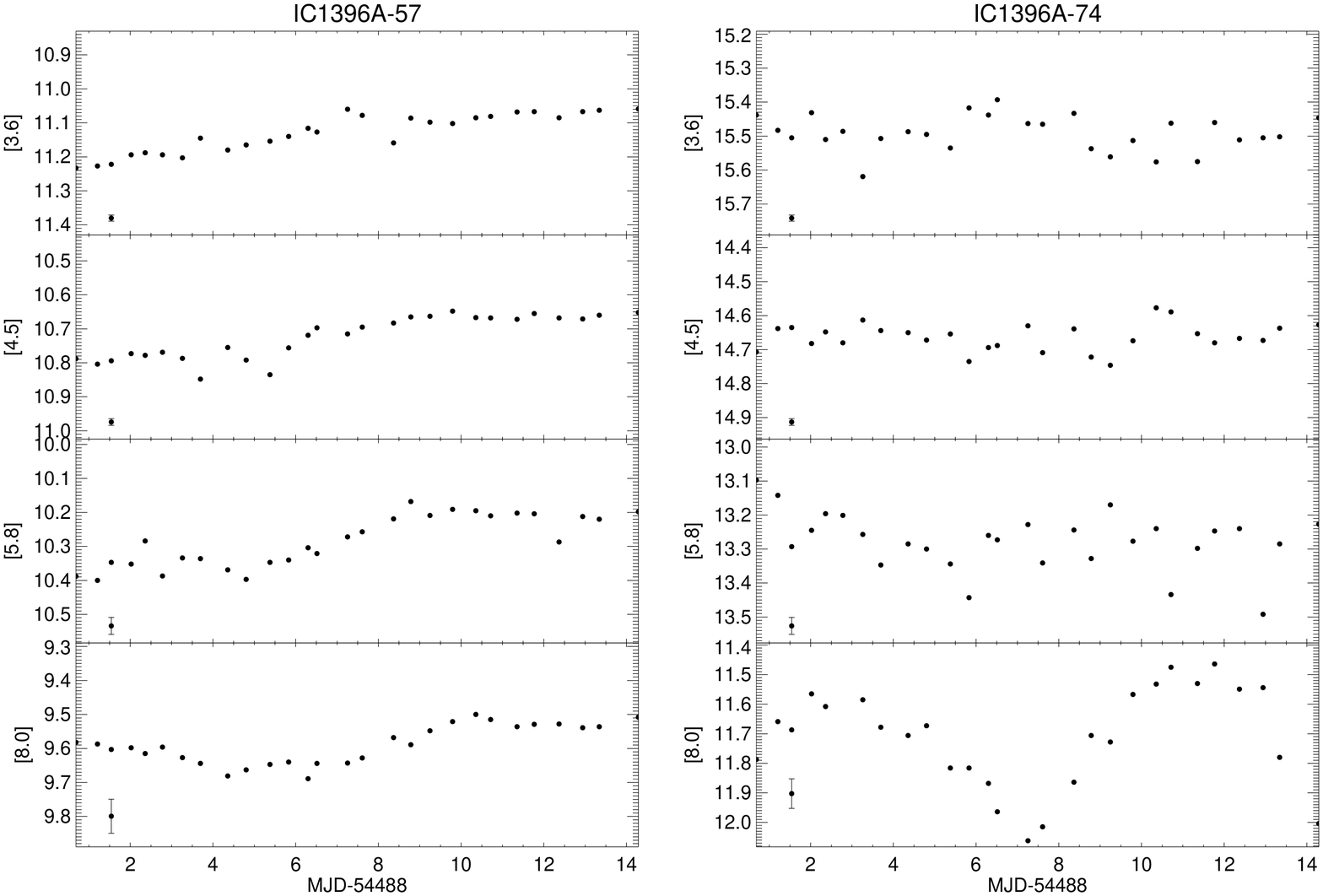}
\caption{{\bf a)}Light curves for IC1396A-57.  This object shows a continuous change in shape from Ch.~1 to Ch.~4. The peak-to-peak amplitude at Ch.~4 is $\sim$0.2 mag. {\bf b)} Light curves for IC1396A-74. The panels, from the top to the bottom, show the IRAC photometry at Ch.~1 (3.6 $\mu$m), Ch.~2 (4.5 $\mu$m), Ch.~3 (5.6 $\mu$m) and Ch.~4 (8.0 $\mu$m) respectively. This object shows almost constant light curves at the three bluer band passes while the star fades in brightness in Ch.~4 for $\sim$6 days. The amplitude of the variation is $\sim$0.5~mag.}
\label{fig:45lc}
\end{figure}

As the other primary means to determine
variability, we also calculated the Stetson J variability index using the four IRAC bands for each source. 12 objects labeled as variable by the $\chi^2$ test were labeled as non variable by the J index however, all but three of them (objects 19, 30, and 51) are labeled as variable if one excludes Ch. 4 to calculate the J statistic. 
We have  given a final tag -- variable (V) or non variable (N)-- to each source based on the $\chi^2$ test, the J index, and a careful visual inspection of all the light curves. 
The very good similarity, in most cases, 
between the light curve shape for Ch.~1 and for Ch.~2 (and 3 and 4 when 
the S/N is good enough) is good evidence 
that the variability is real.    Only one star showed what we believe is
real variability in Ch.~3 or 4 but not in the shorter wavelength channels. 
That object is IC1396A-74 (see later discussion and Figure~\ref{fig:45lc}b).

Table~\ref{tab:longvar} presents 
the results of all targets including IRAC magnitudes, rms amplitudes, 
result of the $\chi^2$ test, J statistic, our final variability tag,
and periods for Ch.~1 \& Ch.~2. Note the use of the term T$_{var}$ instead of rotation period in the table indicating that these are variability timescales rather
than periods since the timespan of our observations are not long enough to confirm the periodicity. Objects are named following the order in Table~6 of \citet{Sicilia06} until number 57. Objects 58 and 59 are YSOs from \citet{Reach04} not in common with \citet{Sicilia06}, and objects from 60 to 74 are our 15 new candidates. YSO classifications for each object  
derived from
the IRAC color-color diagram, are also 
included as well as the YSO classification previously given for these objects. There are 8 objects marked as having different 
classification from that in \citet{Sicilia06}. 
Four of these 
 as Class~I by one of us, and Class~II by the other,  but are really very close to the 
boundary between those classes. Actually two of those objects (1 and 40) 
are in common with \citet{Reach04} where they were classified as Class~I/II. 
The remaining 4 objects were classified by us  as Class~III because of the lack of IR excess. One of them (-30, 
previously known as LKHa 349a) was also studied by \citet{Reach04} and 
classified as Class~III. The remaining 3 objects could be field stars (two of them, 9 and 32 were classified 
by \citet{Sicilia06} as possible non members).

\begin{figure*}[]
\epsscale{0.9}
\plotone{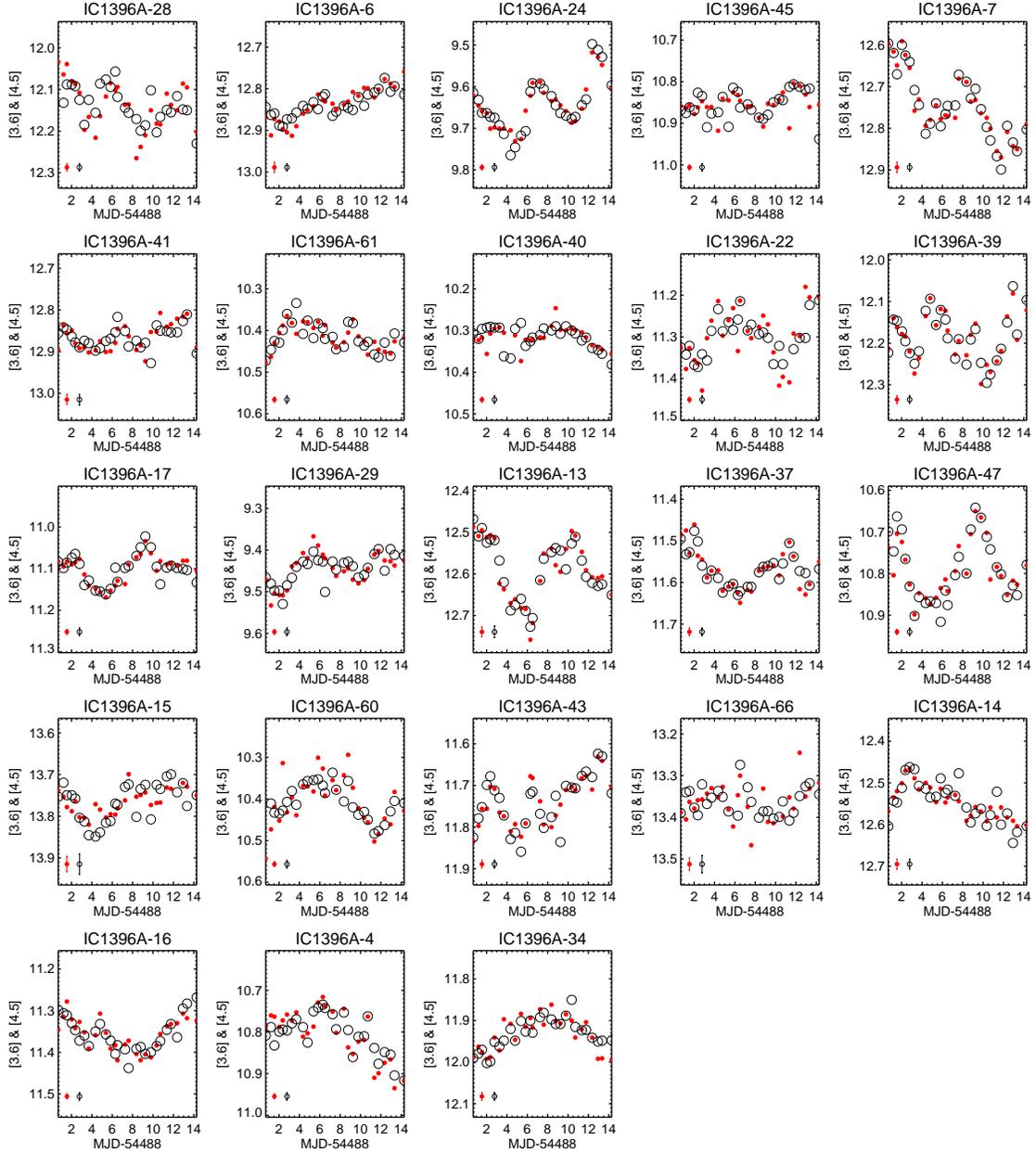}
\caption{Light curves for the 18 IC1396A YSOs which present periodic-like, colorless variations. In the Figure we show objects that seem to have increasing periods (from 5 to more than 14 days) from the top left panel to the right bottom one. Red filled circles represent the Ch.~1 data while open black circles show the Ch.~2 data. The Ch.~2 light curves have been shifted in the y axis to match the mean magnitudes in Ch.~1.
}
\label{fig:19lc}
\end{figure*}

\begin{figure}[]
\epsscale{1.}
\plotone{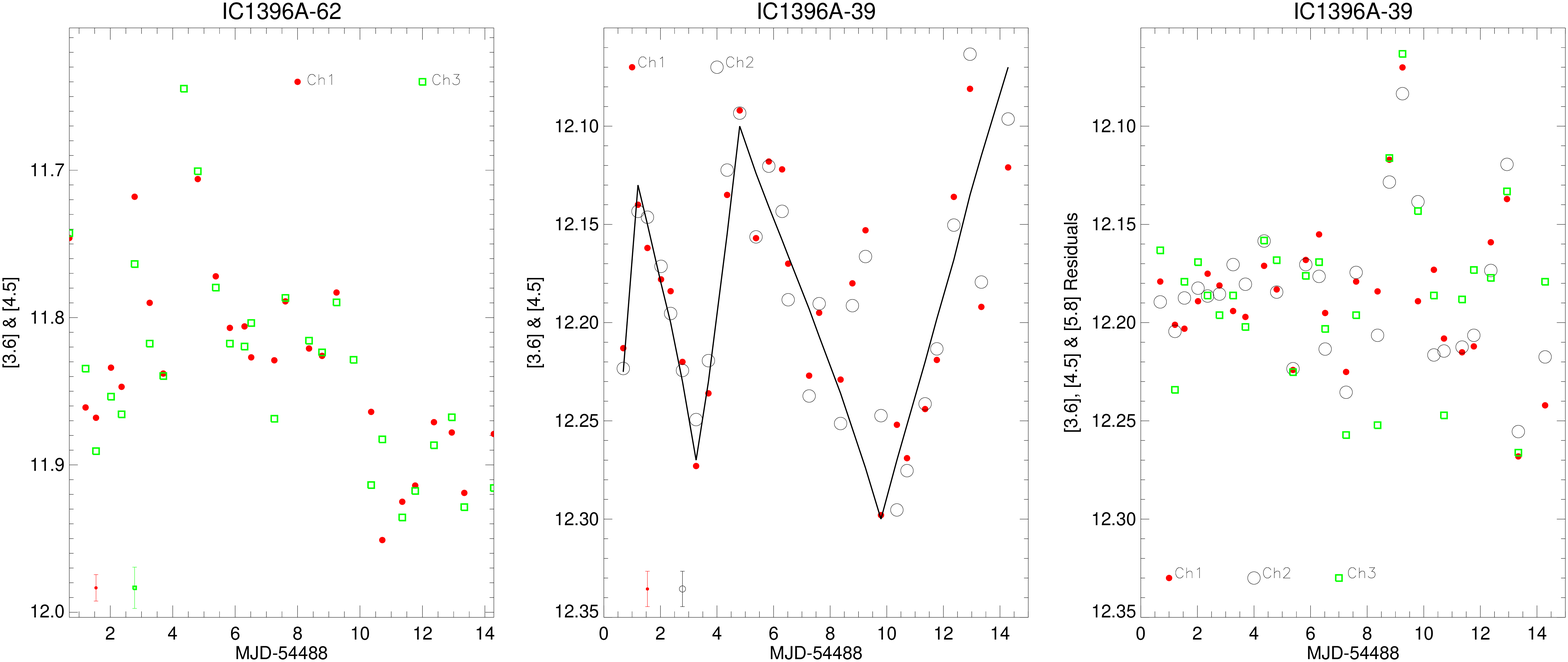}
\caption{{\bf a)} Light curves for IC1396A-62.  This object shows a sudden change in brightness of about 0.2 mag that lasts for 2 days. The object is located just at the border of Ch.~2 and Ch.~4 FOV and thus the light curves presented are those for Ch.~1 (filled circles) and Ch.~3 (open  squares). {\bf b)} Light curves for  IC1396A-39, Ch. 1(red dots), Ch. 2 (black circles), with a saw-tooth pattern illustrating the long-term variation. {\bf c)} Residuals from the light curve in (b) after subtracting the saw-tooth pattern showing  a flare or accretion event. Ch.~3 data is also included (open squares).}
\label{fig:58lc}
\end{figure}

\begin{figure}[]
\epsscale{1.}
\plotone{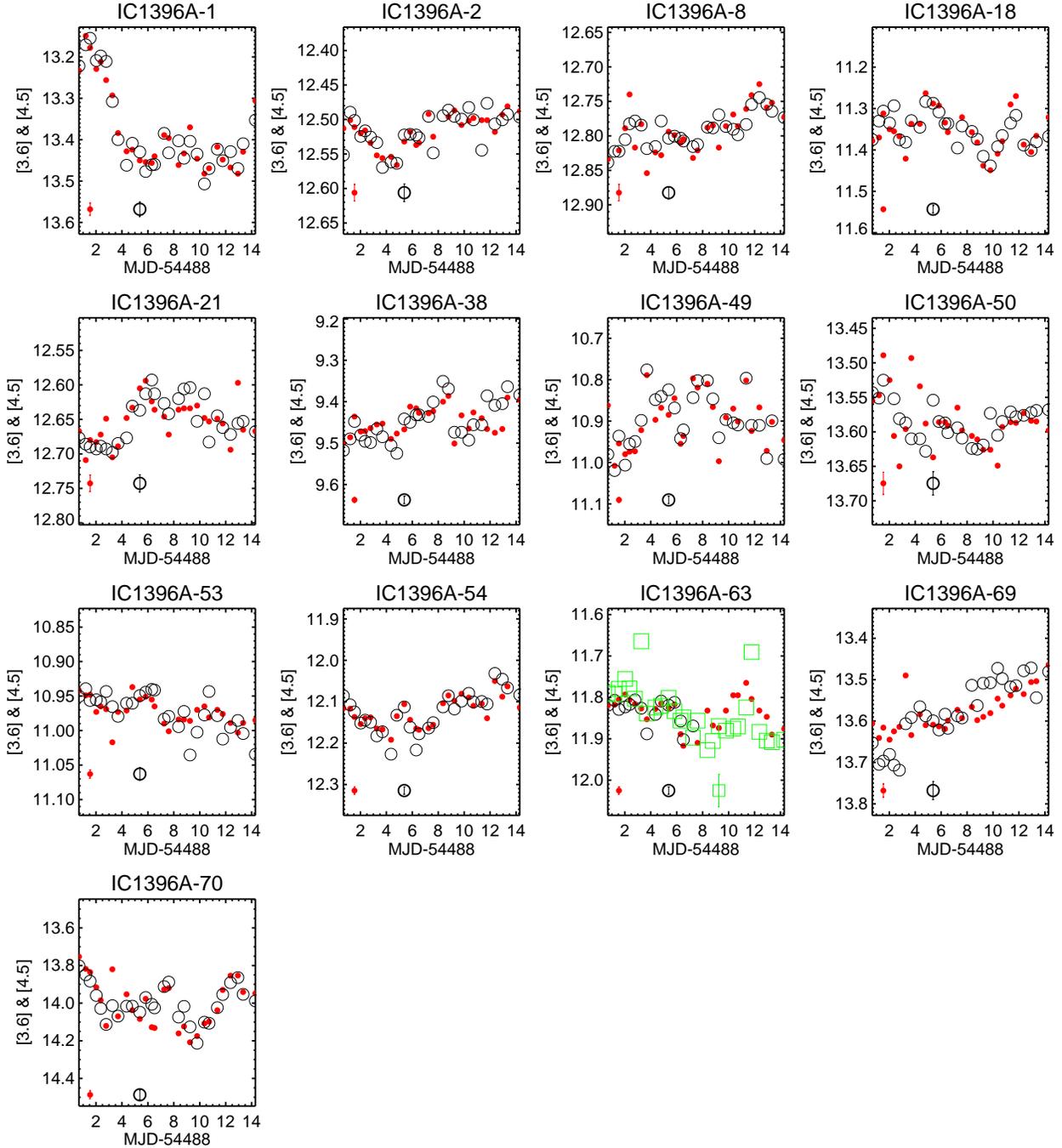}
\caption{Light curves for the 18 IC1396A YSOs which show slow, non-periodic colorless variability. In the figure Ch. 1 (red filled circles) and Ch. 2 are plotted (black open circles). For object 63 the Ch. 3 data -- green open squares-- is also plotted.}
\label{fig:14lc}
\end{figure}

\begin{figure*}[]
\epsscale{1.}
\plotone{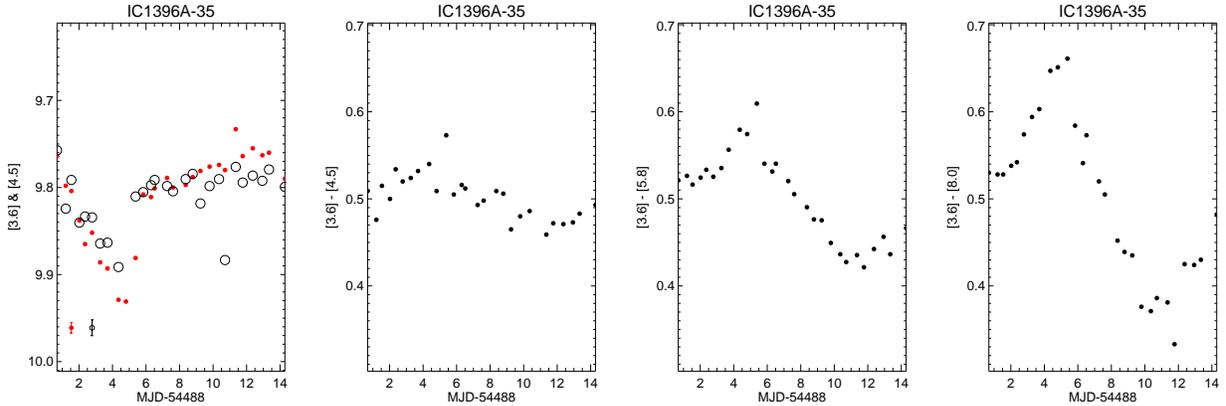}
\caption{Light curves for IC1396A-35. Magnitude change with time in the rightmost diagram and changes of colors with time in the remaining diagrams.  The star presents a decrease in brightness at the beginning of the observation, probably due to some obscuration process, and 4 days later the light curve becomes constant. The peak-to-peak amplitude in the color variation increases from 0.05 mag in [3.6]-[4.5] to 0.3~mag in [3.6]-[8.0].
}
\label{fig:28lc}
\end{figure*}

There is a range of shapes found in the light curves and 
Figures~\ref{fig:19lc}-\ref{fig:28lc} present the different observed 
mid-infrared variability characteristics. All variable stars have their photometry tabulated in Table~\ref{tab:timeseries}, available in the electronic
version of this article.
About 30\% of our YSO's show light curves that resemble the optical
variations of  BY Dra variables (spotted stars whose apparent
luminosities vary due to rotational modulation -- \citet{Byrne87}).  T Tauri stars also
show such variability in the optical and near-IR, 
generally attributed to either
cool, solar-type spots or hot spots associated with accretion flows
onto the stellar photosphere \citep{Rydgren83,Bouvier86,Vrba86}.  It is not obvious, however,
whether our mid-IR light curves can be attributed to such spots (see discussion
in Sec.~\ref{variabilitycauses}).
Figure~\ref{fig:19lc} shows these periodic-like light curves, 
where the characteristic timescale of variability increases from the upper 
left panel to the lower right one. The peak-to-peak variability amplitudes of these 
objects can be found in Table~\ref{tab:amp}.  These data illustrate that 
the variability amplitudes are essentially invariant with wavelength from 
3.6 to 8~$\mu$m. 

Other types of variability exhibited by one or more of our YSO's
include:
\begin{itemize}

\item{rapid (timescales of hours) variability inconsistent with rotational
modulation.  No apparent color dependence during the variations.  
Only two objects present this kind of variations: IC1396A-39 (object $\eta$ in \citet{Reach04}) and IC1396A-62 -- see Figure~\ref{fig:58lc}.
Objects 39  and 62 are Class~I and Class~II, respectively.
Possible causes:  flares or accretion flickering;}

\item{slow (timescales of days), non-periodic variability changes, with little
or no color changes.  20\% of our sample of YSOs fall under this catergory. Their light curves are shown in Figure~\ref{fig:14lc}.
Only three Class I YSOs (IC1396A-1, IC1396A-8, and IC1396A-50) are included in this group. Possible causes:  accretion variations; rapidly evolving spots. We admit there is in some cases ambiguity whether
an object should be in Fig~\ref{fig:19lc} or~\ref{fig:14lc} - the time span of our observations
is  too short for certainty.}

\item{slow (timescales of days), flux modulation with color dependence.
Three objects of our sample show this kind of variations:  IC1396A-57 (Figure~\ref{fig:45lc}a), which shows a slow 
brightening over the entire 14 days of observation with a slower rise
to maximum going to longer wavelengths, and IC1396A-35 (Object $\beta$ in \citet{Reach04} --Figure~\ref{fig:28lc}),
which shows a slow fading over about five days, followed by a sudden
reset to the original brightness and then approximately constant flux
at 3.6 and 4.5 microns for the remaining time.  IC1396A-35 is the only
YSO with significant, coherent color variation over the 14 day observing
run - becoming significantly redder when faint.  Another possible
exemplar of this class is IC1396A-74, which shows essentially constant
magnitude at 3.6, 4.5 and 5.8 microns, but an apparently significant
decrease in flux over a several day period at 8 microns (Figure~\ref{fig:45lc}b).  
Objects 35 and 57 are Class~II stars and object 74 is a Class~I YSO.
Possible causes: radially differential heating of the inner disk; obscuration by an inner disk
over-density. }
\end{itemize}

\subsection{The staring data: short-term variable members.}\label{Sorttermvariability}
%
%

\begin{figure*}[]
\centering
\includegraphics[angle=0,scale=0.2]{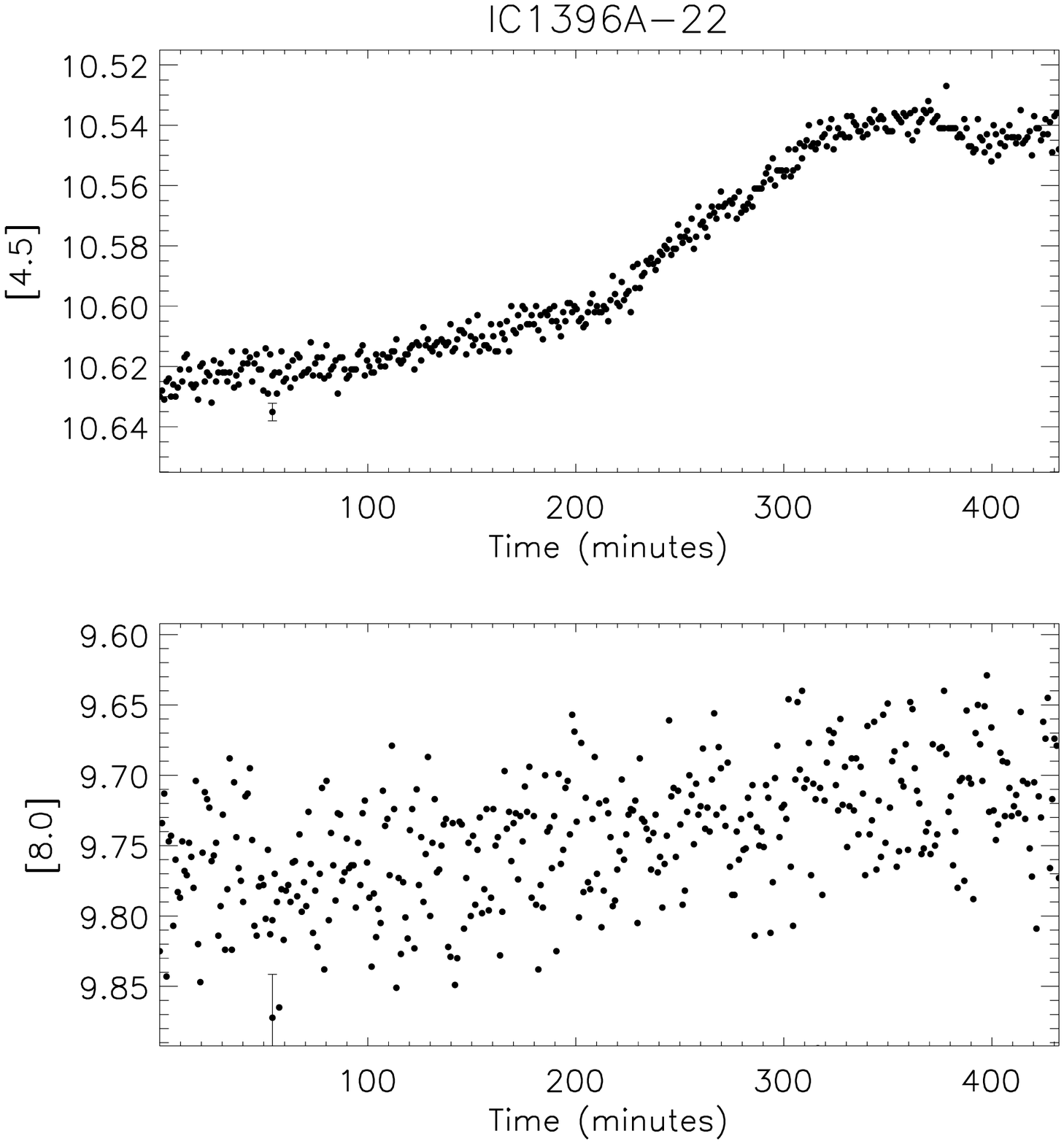}
\includegraphics[angle=0,scale=0.2]{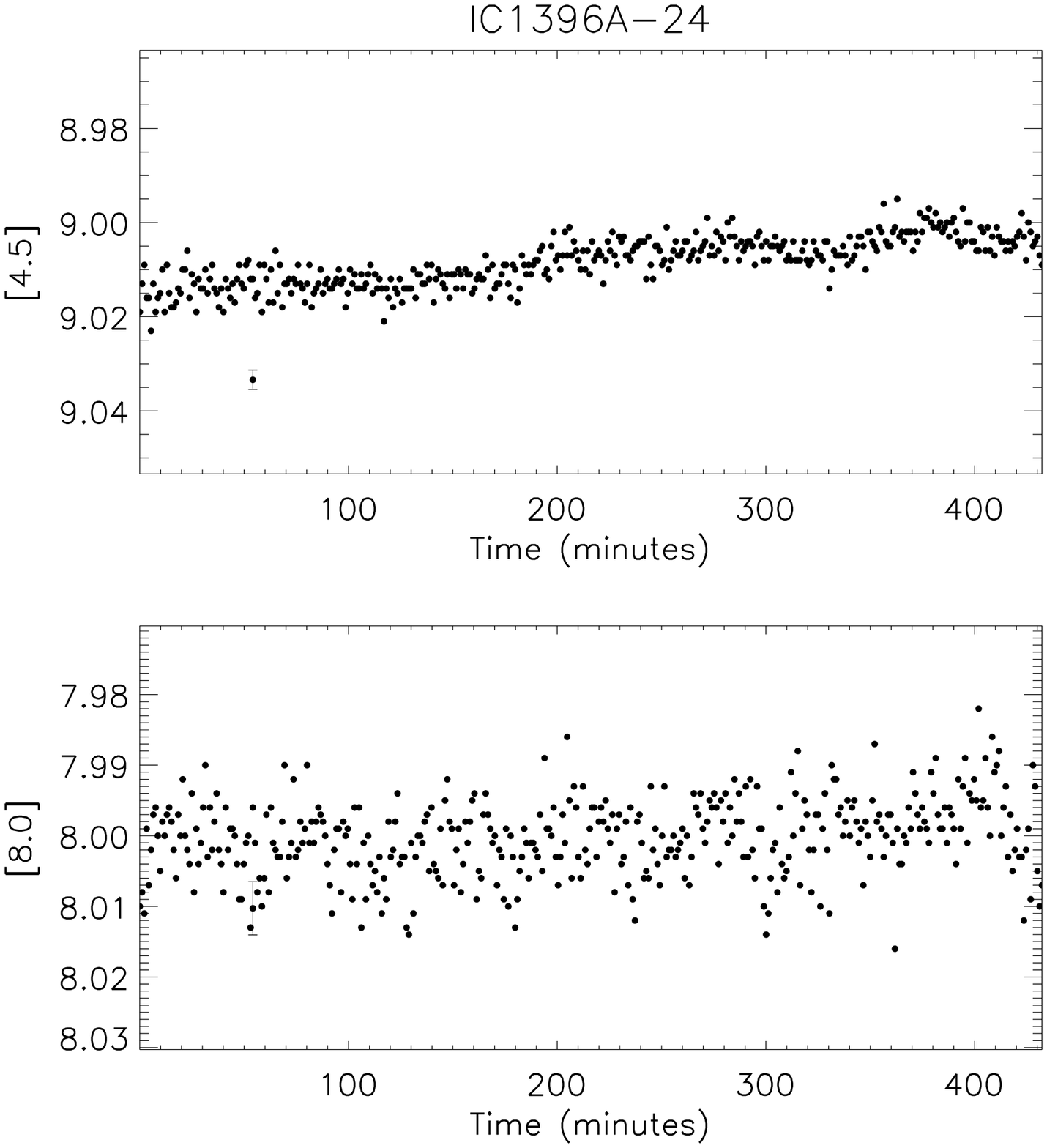}
\includegraphics[angle=0,scale=0.2]{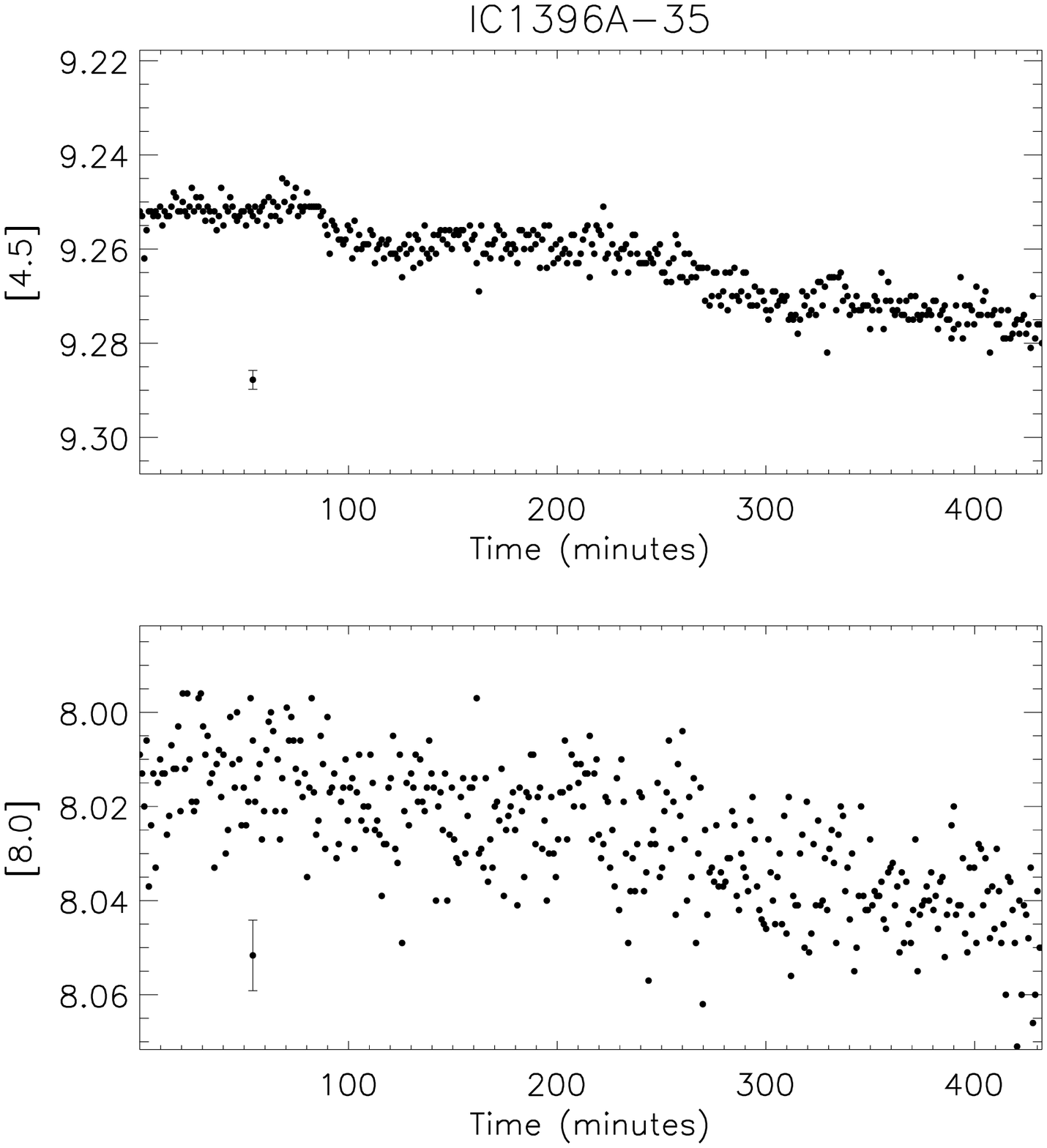}
\includegraphics[angle=0,scale=0.2]{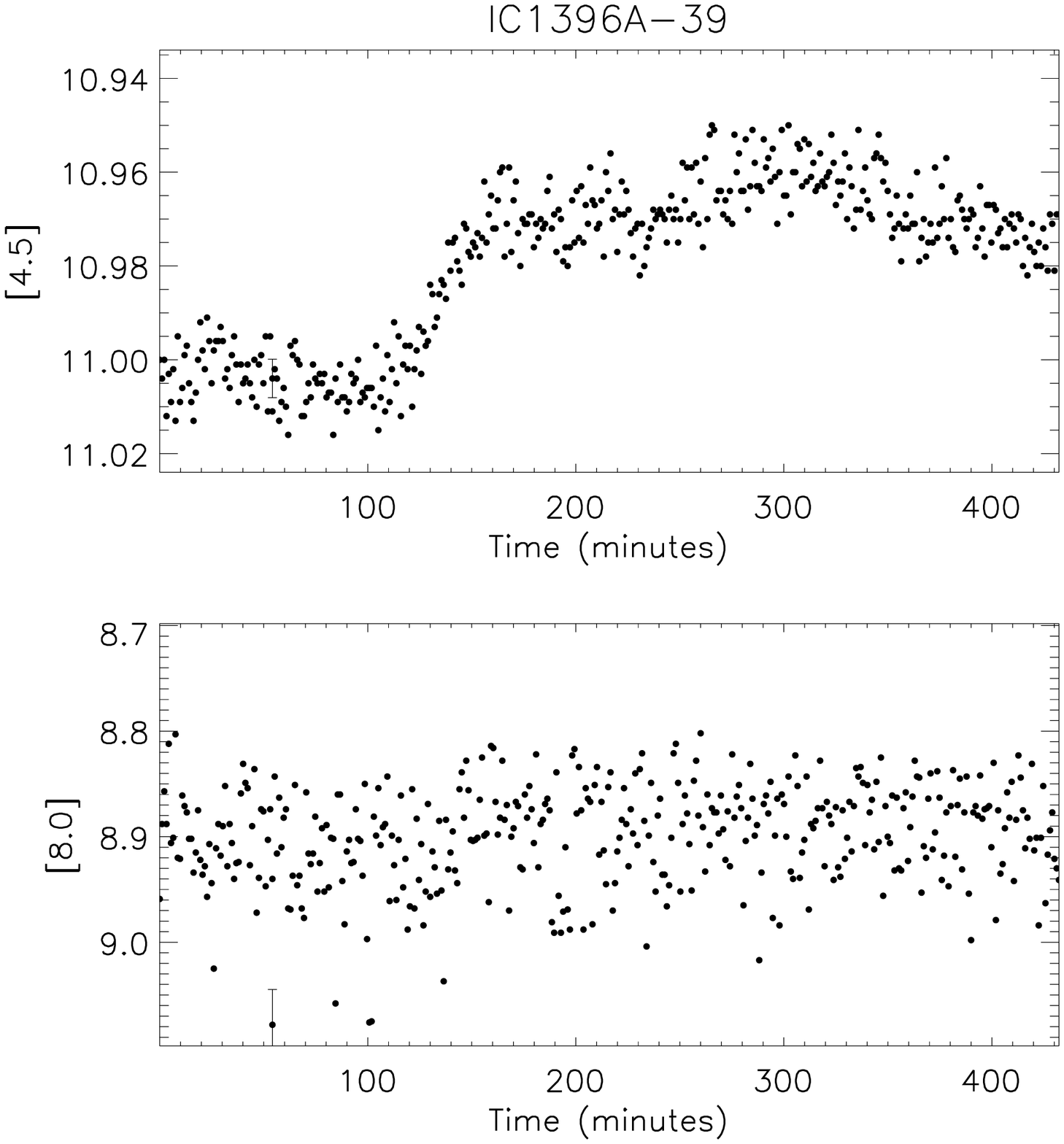}
\includegraphics[angle=0,scale=0.2]{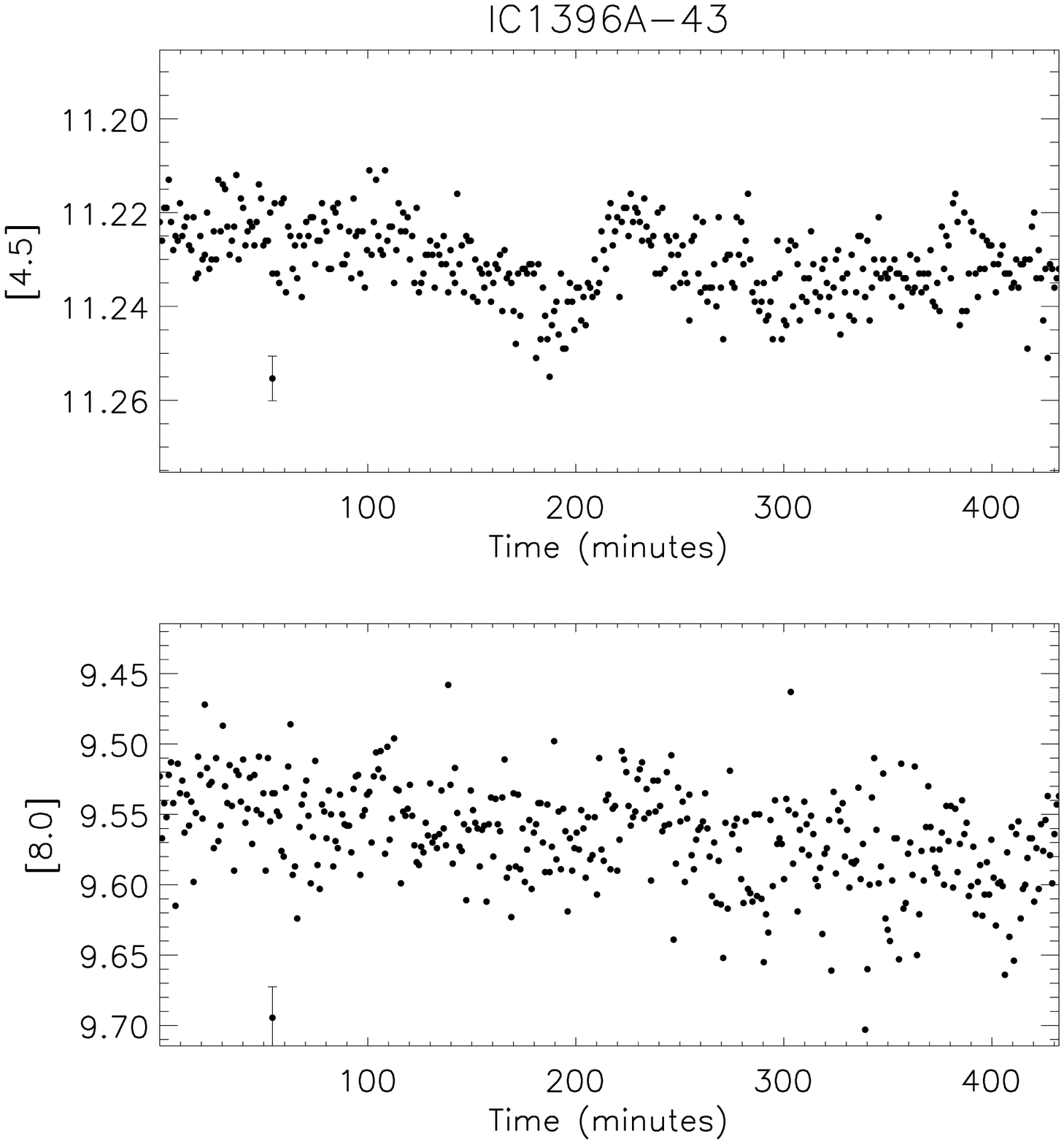}
\includegraphics[angle=0,scale=0.2]{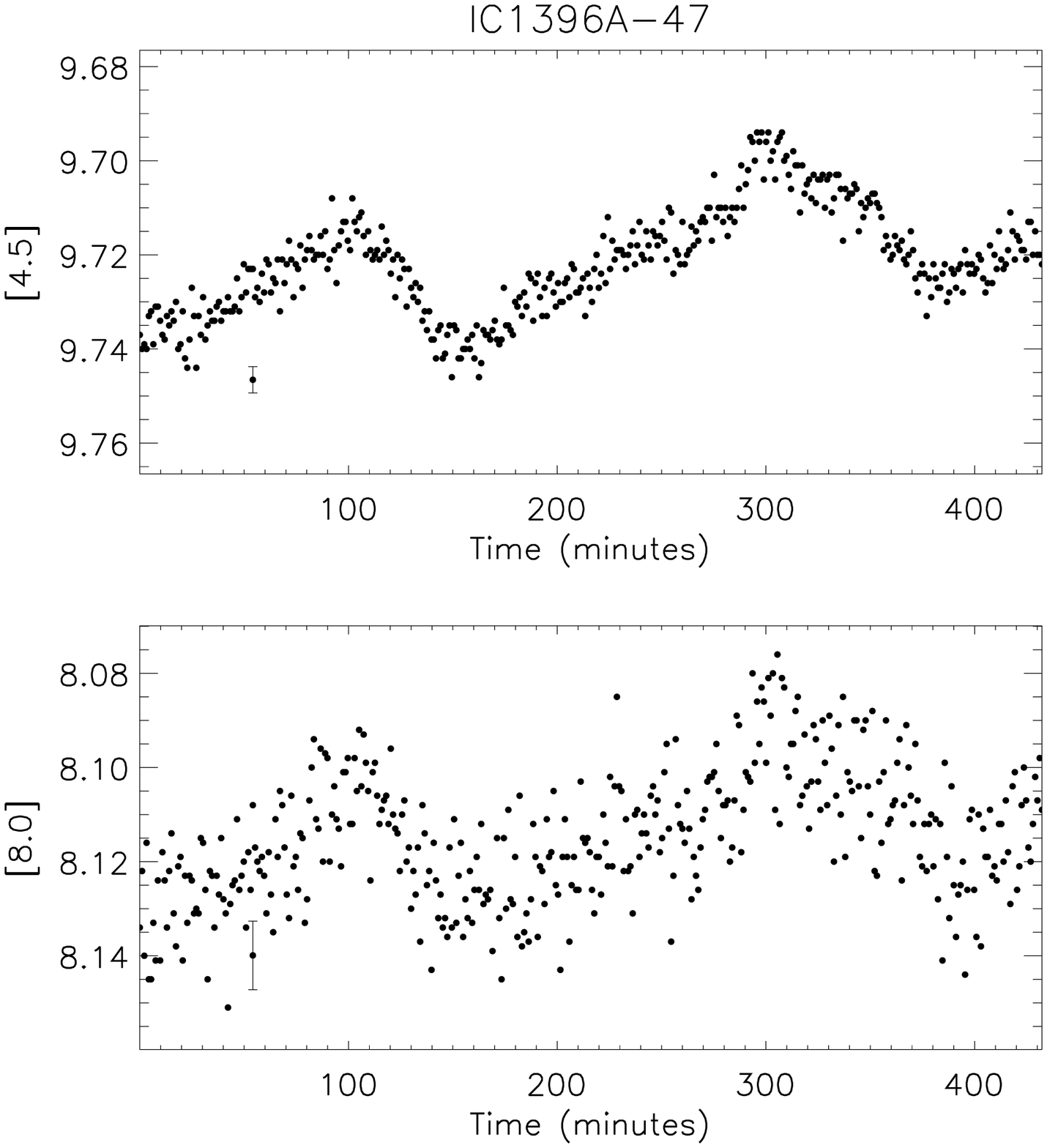}
\includegraphics[angle=0,scale=0.2]{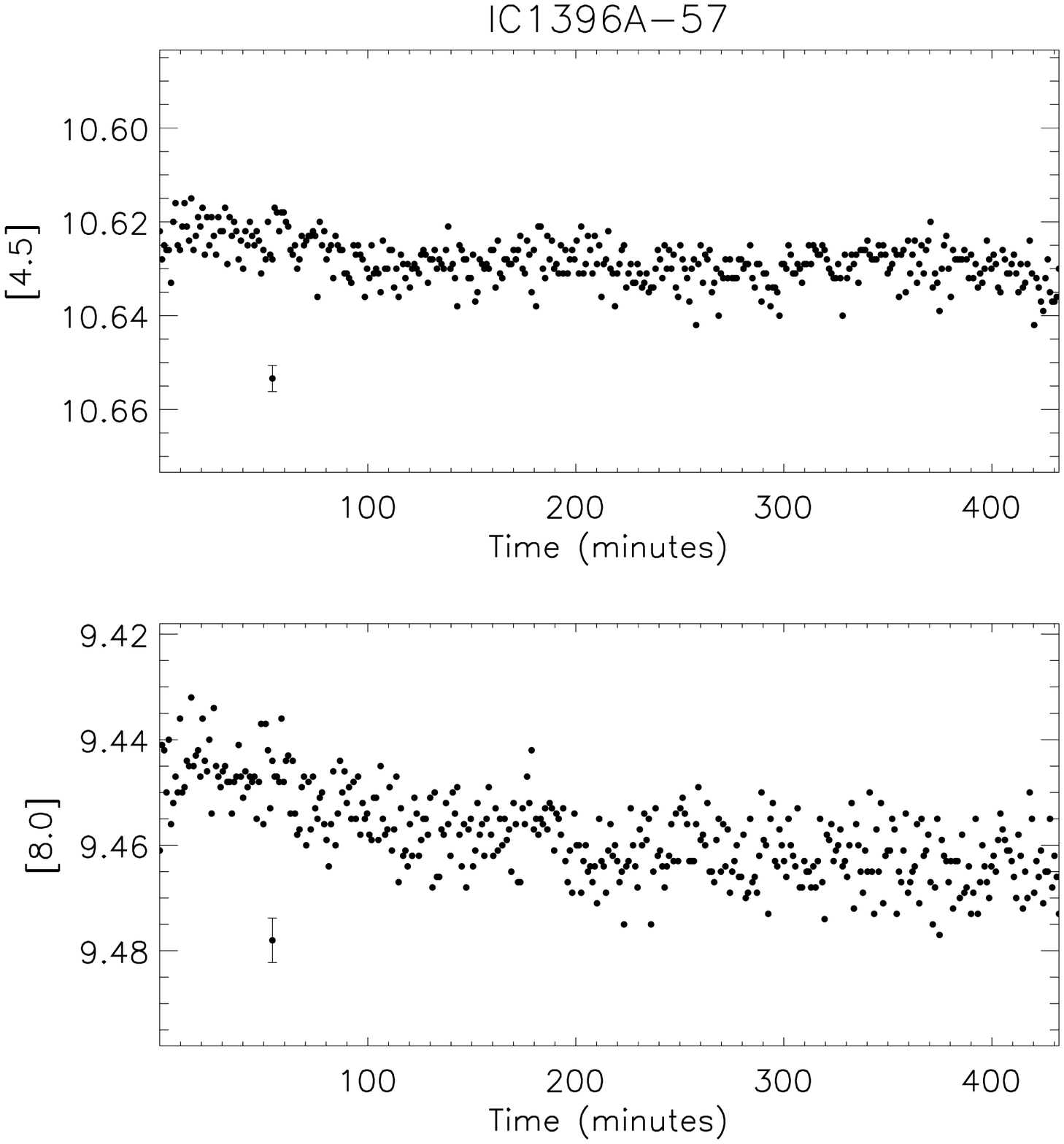}
\includegraphics[angle=0,scale=0.2]{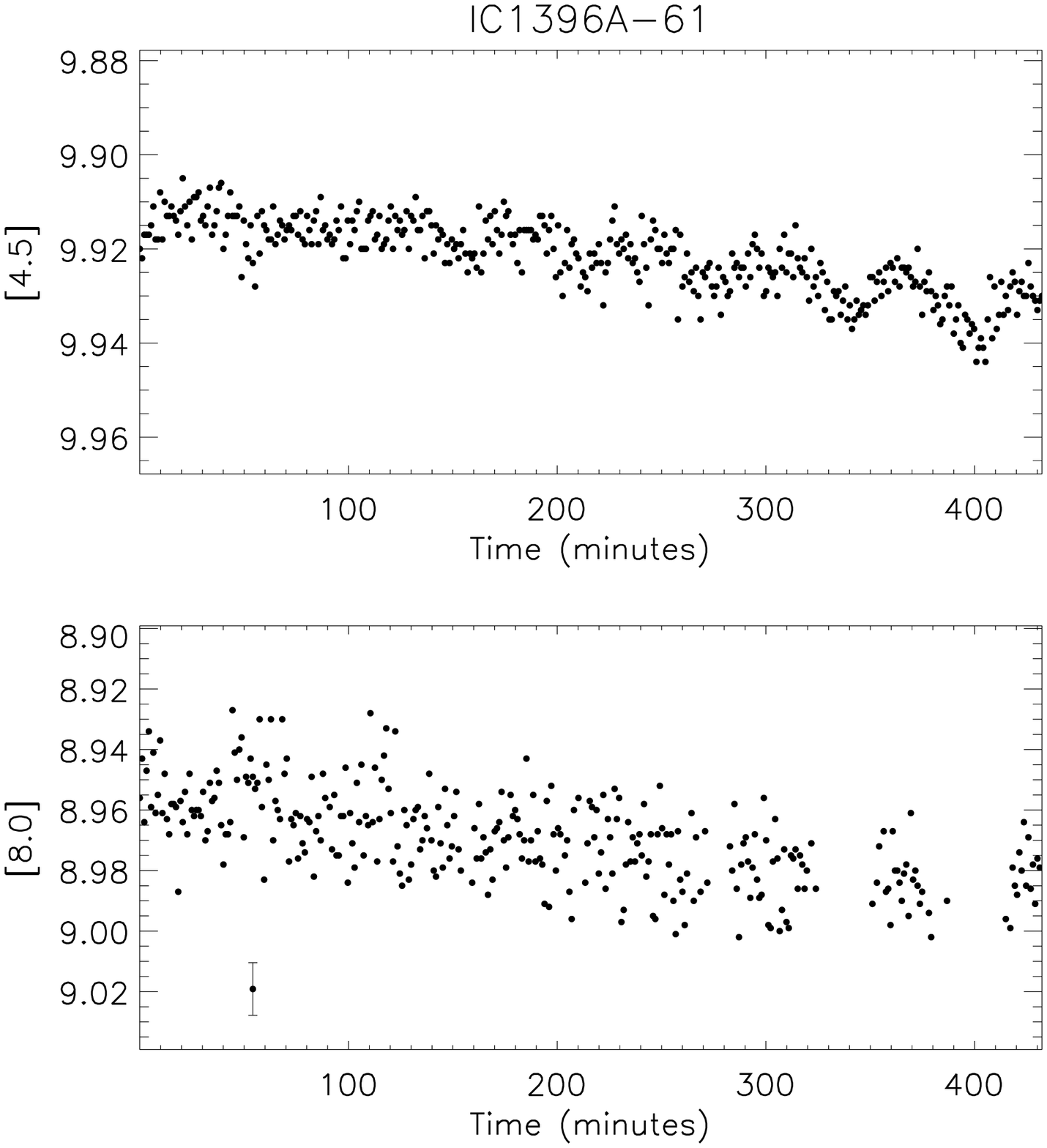}
\caption{Light curves for the eight IC~1393A short term variable YSOs. The top and bottom panels show Ch. 2 and Ch. 4 time series, respectively. The rms uncertainty of a single point is represented in the lower left corner of each panel. Note the different scales in the y axis in the  upper and lower panels.
}
\label{fig:shortvar}
\end{figure*}

\begin{figure}[]
\epsscale{0.4}
\plotone{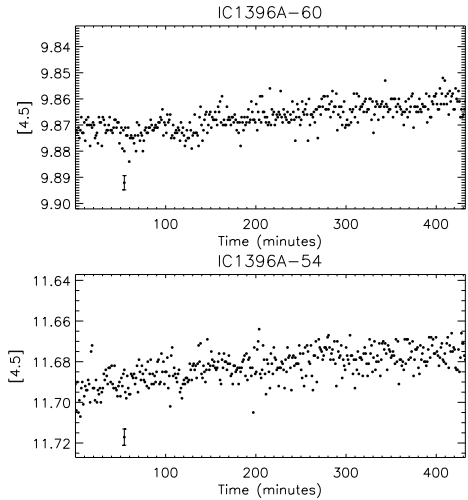}
\caption{Ch. 2 light curves for the two IC~1393A YSOs which lacking the Ch. 4 light curves, do show an upward trend in their Ch. 2 light curves. The rms uncertainty of a single point is represented in the lower left corner of each panel. 
}
\label{fig:shortvar2}
\end{figure}

Thirty eight out of the 42 targets that fall inside the FOV of Ch.~2 \& Ch.~4 
in the staring mode observations have 
enough good photometry to derive their time series, at least in one of the 
two band passes. The remaining objects  were 
saturated in our images (IC1396A-19 and IC1396A-30) or were too faint to obtain their photometry (IC1396A-26 and IC1396A-27). Five out of these 38 objects are new candidate members 
based on our new deep photometry. Table~\ref{tab:shortvar} presents the 
results of all targets including IRAC magnitudes, rms amplitudes for the 
light curves, result of the $\chi^2$ test, the period of the modulation observed if there is any, the J variability index, and our final tag of variability. 14 objects were 
labeled by the $\chi^2$ statistic as variable in Ch.~2, however only 4 of 
them were labeled as variable in both band passes, IC1396A-35, IC1396A-47, IC1396A-57, 
and IC1396A-61 (Figure~\ref{fig:shortvar} shows the light curves of the short-term variable stars). Another 7 objects where labeled as variable by the J variability index.
The most interesting object is IC1396A-47, which shows a period of 3.4~hr in both band 
passes (we discuss this star in more detail in Sec.~\ref{zeta}).
A few of the 
objects labeled as variable in Ch.~2, but not in Ch.~4 by the $\chi^2$ test, have light curves 
in Ch.~4 which follow the same trend and have similar amplitude to that of the light 
curve in Ch.~2. They have not been labeled as variable at Ch.~4, probably 
due to a much higher noise at 8.0~$\mu$m than at 4.5~$\mu$m. These objects are IC1396A-22, IC1396A-24, IC1396A-39, and IC1396A-43 -- two of them are variable based on the J index and the other two are just below the adopted J index limit for variable stars. We consider the four of them as variable objects due to the shape similarity between both bandpasses. The 
remaining objects show amplitudes in Ch.~2 of order of the noise seen in 
the light curves in Ch.~4, not allowing the detection of variability in 
the latter channel. There is only one object, IC1396A-25, which has detected 
variability based on the $\chi^2$ test only in Ch.~4. This object is variable based on the J index however, it is very bright at 8~$\mu$m and since 
latent build-up seems to be flux dependent, it is possible that our 
correction is not good enough for this object and the variability is instead 
spurious. The four objects with detected variability in both bandpasses 
show the same light curve shape in each band, IC1396A-35 and 
IC1326A-61 show a downward trend during the 7.5~hr of observation,  and IC1396A-47 
presents a pulsation like variation (see Sec.~\ref{zeta}). The last object labeled as variable in both channels, IC1396A-57, shows
$\chi^2$ values that are only slightly above our detection threshold. Its light curves show a low amplitude correlated variation and it is labeled as variable by the J index.  IC1396A-57 does vary on long timescales, as shown in Figure~\ref{fig:45lc}.
Finally there are two objects which lack good photometry in Ch. 4 --IC1396A-54 and IC1396A-60-- but which show an upward trend in their Ch. 2 light curves. Since we cannot confirm the variability with the Ch. 4 data we do not include those objects in the final list of variable YSOs. Their light curves can be seen in Figure~\ref{fig:shortvar2}.

To summarize, 8 out of 38 YSOs show correlated short-term variability in Ch. 2 and Ch. 4. 
All these objects also show long-term variability (see Sec.~\ref{longtermvariability}) 
and the variations seen in the 7.5~hr period are consistent with the
light curve shapes seen in the mapping mode data 
(note that the staring observations were performed 
5~hr after the last mapping observation was finished). 

The remaining objects are catalogued as non-variable in short-timescales and any possible 
variability has to be under the rms amplitudes quoted in Table~\ref{tab:shortvar}.

\subsection{Variability of stars without IR excesses in the mapping data}\label{nonmembers}
%
%

\begin{figure}[]
\epsscale{0.6}
\plotone{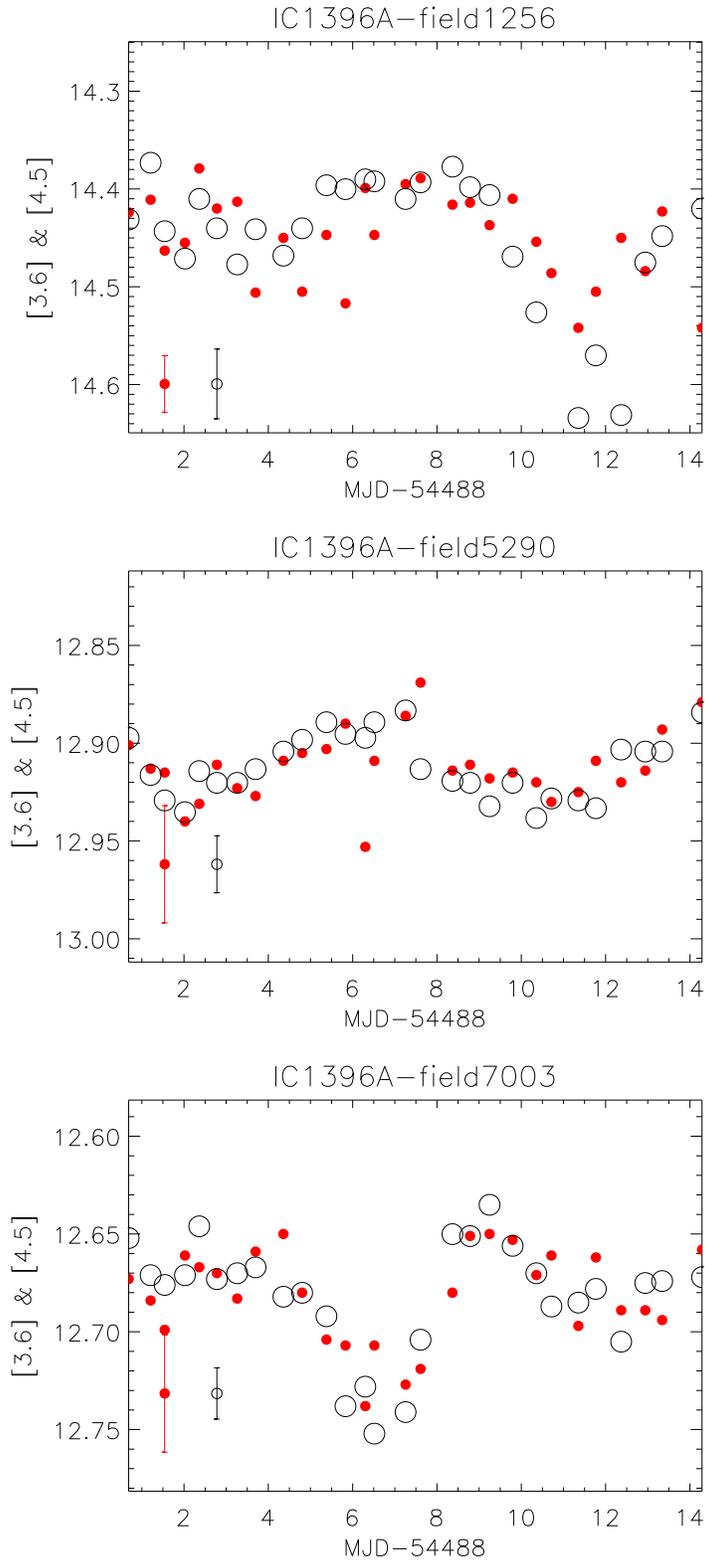}
\caption{Light curves for a few variable candidate members of IC1396A from IRAC time-series observations. Red solid circles stand for 
the IRAC photometry at Ch.~1 (3.6 $\mu$m) while black open circles represent IRAC photometry at Ch.~2 (4.5 $\mu$m).  The 4.5 $\mu$m light curve has
been shifted in the y direction so that the mean magnitude is the same as for the 3.6 $\mu$m points.
Ch.~3 \&  Ch.~4 light curves (not shown) show the same trends as in this plot but they are noisier.
}
\label{fig:longvarnonmem}
\end{figure}

In addition to the previously known members we have inspected the 
``field" (which includes both true field stars and YSO's lacking
IR excesses)
stars, in search of additional variable stars. Due to their lack of infrared excess, 
Class~III YSO's cannot be separated from field objects at {\em Spitzer} 
wavelengths. However, given that photometric variability is one of the original 
defining characteristics of pre-main sequence stars, the 
detection of variability can be used as a possible indicator of youth. Thus, 
we applied the $\chi^2$ and J statistics to the entire sample of detected objects 
in Ch.~1 and Ch.~2. 6$\%$ of the sample was labeled as variable in both tests. We visually inspected the light curves of all these objects 
to remove spurious variability detections due to bad photometric measurements 
or very noisy light curves. We identify 46 objects as being variable 
both in the visual inspection and in the statistical tests. Most of the light curves of these objects are 
consistent with rotational modulation with periods of order 4- $>$10 days. 
These objects present Rayleigh-Jeans-like SEDs with 
no infrared excess other than contamination expected from the nebulosity. 
While we are reasonably convinced these stars are variable at 3.6 and 4.5 microns, 
without further data we cannot know whether they are members of 
IC1396A, members of the older Tr37 cluster, spotted field stars, or other 
variables of uncertain type. Data at other wavelengths are needed to
categorize these stars. Light curves for a few of these new candidates 
can be seen in Figure~\ref{fig:longvarnonmem} and their photometry is presented in Table~\ref{tab:fieldvariables}.

\subsection{Properties of the variable YSOs}\label{properties}
%
%

As summarized in Table~\ref{tab:statistics}, 41 out of the 69 objects that 
form our sample of YSOs (54 previously known YSOs plus 15 new candidates) are variable on long timescales.
Forty percent of the 69 YSOs in Table 2 show 
peak-to-peak amplitudes of at least 0.1~mag and 60\% of the sample 
present peak-to-peak amplitudes greater than 0.05~mag.  Most of the variable 
YSOs show colorless variations (at least within the photometric errors) but 
there are 3 objects whose IRAC colors vary with time. In two out of 
them the object gets redder as it gets fainter, as would be expected if 
variable extinction or rotational modulation of a spotted photosphere is the cause of the photometric 
changes. 23 YSOs show what seems to be well-defined apparently periodic variability 
with amplitudes up to about 0.2 mag. 
Where we are able to derive estimated periods, the periods range from
about 5 days to about 11 days.
These variables include 9 Class I 
and 14 Class II objects.  

\begin{figure}[]
\epsscale{.5}
\plotone{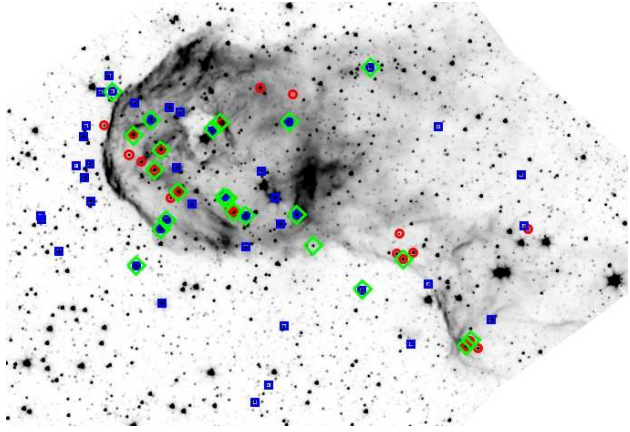}
\caption{Spatial distribution of YSOs in IC1396A. Red circles represent the Class~I objects, blue squares show the Class~II objects, and stars with periodic-looking are plotted with green large diamonds.
}
\label{fig:classes}
\end{figure}

\begin{figure}[]
\epsscale{0.5}
\plotone{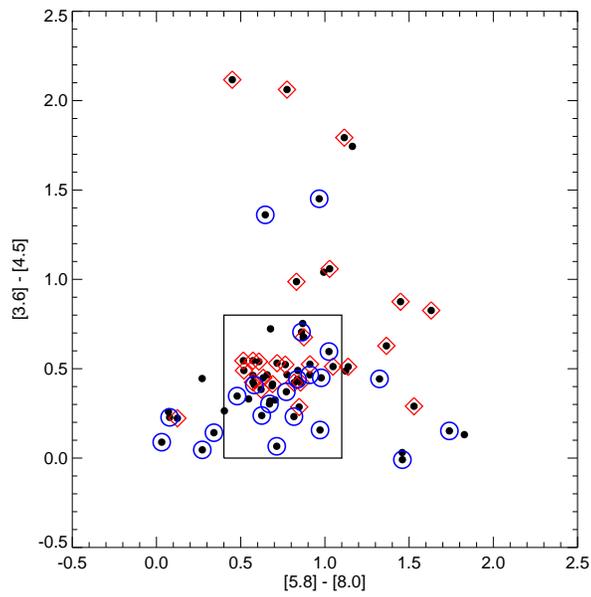}
\caption{IRAC color-color diagram. Red open diamonds represent objects with variability amplitudes larger than 0.1~mag and blue circles represent constant, or almost constant YSOs.
}
\label{fig:iracccdvar}
\end{figure}

In order to help determine the physical cause(s) for this variability,
it is useful to determine if there is anything special about the variable stars
- i.e. are they younger than average? are they more embedded than average?
are they higher or lower mass than average? 
Figure~\ref{fig:classes} shows the spatial distribution of the different types of YSO's. The figure illustrates that all the Class~I and most of the objects with periodic-like light curves are located within the most embedded areas of the nebulosity while the Class~II stars seem to be more spread out across the mosaic. Figure~\ref{fig:iracccdvar} shows the IRAC [3.6]-[4.5] vs. [5.8]-[8.0] 
diagram where the YSOs of different types have been identified (red open diamonds represent large 
amplitude variables --peak-to-peak amplitudes of at least 0.1~mag-- while 
blue open circles stand for constant objects). The diagram shows that the large amplitude variables tend to be redder in the IRAC colors 
(though some other stars are also quite red and do not show significant variability with IRAC). 
We conclude from Figures~\ref{fig:classes} and~\ref{fig:iracccdvar} that while our large amplitude variables
can be either Class I or II, there is a preference for the large amplitude
variables to be amongst the evolutionarily youngest stars in our sample. 
In addition, if we compare the whole sample of YSOs with the group of objects showing periodic-like variations the magnitude distributions are very similar and there is no evidence that the latter ones are more or less massive than the whole sample.

\citet{Carpenter01} presented a near-IR variability study toward a several square degree 
area in the Orion A molecular cloud. They found more than a thousand 
variable stars with a high diversity of photometric behavior. As in our case, 
most of their variable objects did not show color variations and 18\%
of them were periodic. 
They concluded  that most of 
the periodic stars were better explained by cool spot models, primarily
because of the small change in color during the variability cycle. Most of the periodic 
stars were compatible with being WTTs, based on their colors. However,
some aspects of the JHK data seemed incompatible with either cool or hot
spot models.
The small number of objects in our sample, the difference in the sample
construction (in particular, the lack of Class III sources in our sample), and
the different wavelength coverages prevents us from making any direct statistical
comparison with the  \citeauthor{Carpenter01} data.  In addition, the thermal emission
from dust at IRAC wavelengths adds a significant complication which makes it
more difficult for the simplest spot models to fit our data, as we discuss
in the next section.
 
\section{Causes of variability}\label{variabilitycauses}
%
%

\subsection{Accretion variation}\label{accretion}
\begin{figure*}[]
\epsscale{1.1}
\plotone{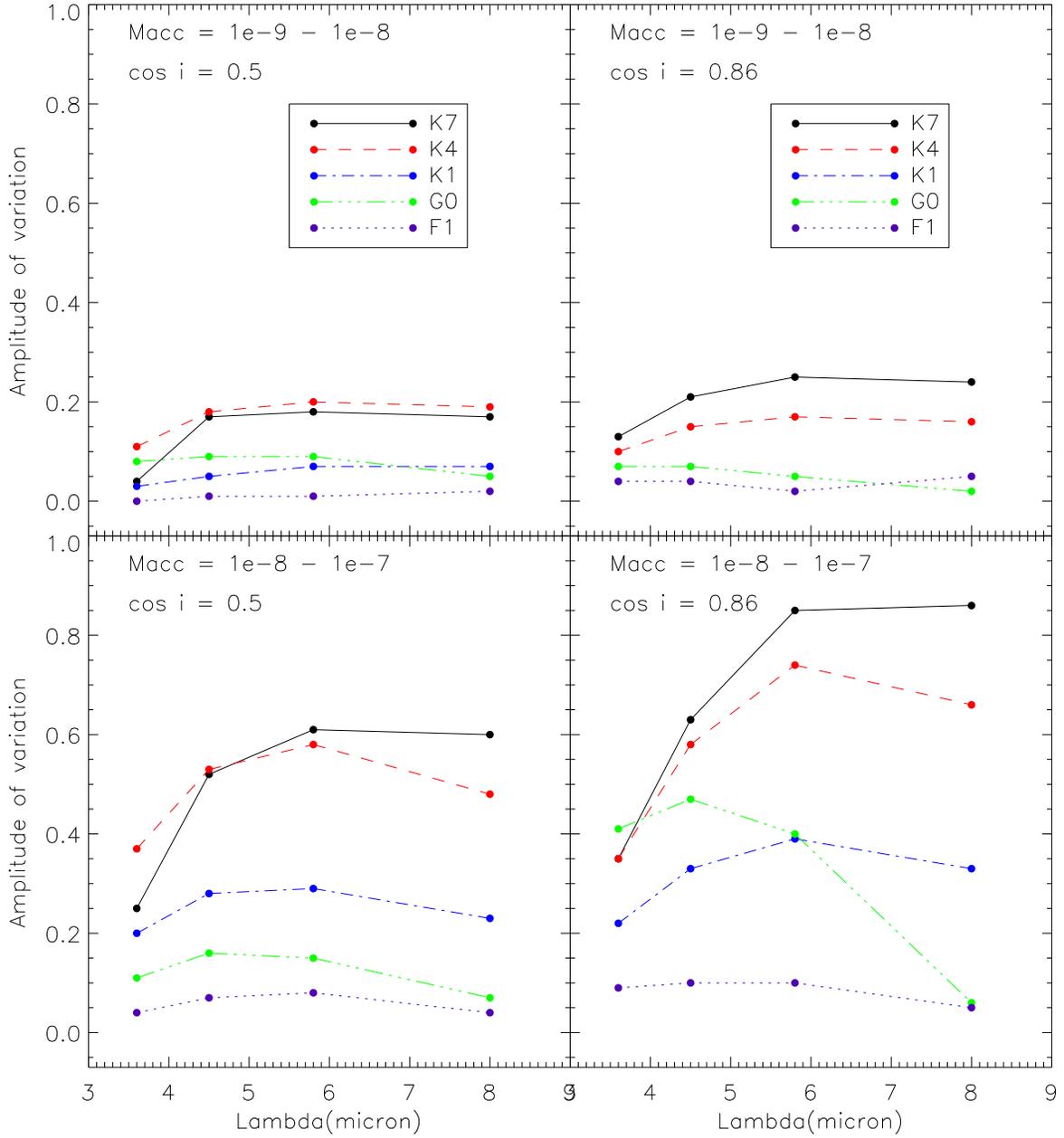}
\caption{Amplitudes predicted by the disk models for a change in the mass accretion rate from $10^{-9}$ M$_\sun$/yr up to $10^{-8}$ M$_\sun$/yr in the upper panels and from $10^{-8}$ M$_\sun$/yr to $10^{-7}$ M$_\sun$/yr in the bottom ones. 
}
\label{fig:dalessio}
\end{figure*}

%
%
We have investigated the possibility that the variation seen in some of our 
light curves come from changes in the disk mass accretion rate. \citet{DAlessio05} 
have constructed a database of accretion disks models. Using these models we 
have computed the amplitudes of the variability due to changes of one order 
of magnitude in the mass accretion rates, from $10^{-9}$~M$_\sun$/yr to $10^{-8}$~M$_\sun$/yr and from $10^{-8}$~M$_\sun$/yr up to 
$10^{-7}$~M$_\sun$/yr. Figure~\ref{fig:dalessio} shows the predicted brightness 
variations computed for two different variations in the accretion rates 
and two different inclination angles. According to these models, one order of magnitude 
change in an accretion rate $\sim10^{-9}$~M$_\sun$/yr could produce the amplitudes we see in our 
light curves particularly if the underlying star is relatively late
spectral type. At least for some parameter ranges, the models also predict
relatively little variation in IRAC color as a result of the change in
accretion rate, as necessary to match most of our variable objects. This mechanism can be causing the non-periodic variability
   changes on multi-day to week timescales seen in some of our variable YSOs. However, we believe that it is not the predominant mechanism; such rapid, extreme
accretion variability would be easily detected at shorter wavelengths, yet
few examples have been reported.

\subsection{Starspots}\label{accretion}
%
%
More than half of the IC1396A variable YSOs have light curves that appear periodic, as shown in Fig~\ref{fig:19lc},  with amplitudes 0.1-0.2 mag 
(See Table~\ref{tab:amp}), and typical periods in the range 5 to more than 
11 days. 
For T Tauri stars observed in
the optical or near-IR such variability is thought 
to originate mainly from either cool magnetic spots (dynamo driven activity) 
or hot accretion spots on the stellar surface (where the material from the disk falls onto the star) 
 that are hundreds to thousands of kelvins different in 
temperature from the photosphere and rotate with the star. Light curve 
amplitudes in the optical of order 0.2~mag (or considerably larger) are not 
atypical for CTTs or WTTs. We have contemplated the possibility that hot 
or cool spots on the photospheres of these YSOs could produce the light 
curves we observe; using somewhat extreme parameters, light curve amplitudes 
of order 0.2 mag can be obtained even at IRAC wavelengths.  We have estimated 
the photometric amplitudes expected from both cool and hot star spots using 
a simple model \citep{Vrba86} which assumes that the stellar surface has 
a spotted region characterized by a single temperature blackbody and is 
confined to one hemisphere.
The amplitude of the light variations as a function of wavelength can be expressed as:

\begin{equation}
m(\lambda)=-2.5 log \{1 -f [1 - B_{\lambda}(T_{spot})/B_{\lambda}(T_{*}]\}
\end{equation}
were $T_{spot}$ and  $T_{*}$ are the temperatures of the spot and the 
photosphere, $f$ is the maximum fraction of the stellar photosphere covered 
by spots and $B_{\lambda}(T)$ is the Planck function. Note that this star 
spot model is an approximation since it ignores limb darkening, inclination
effects, and opacity differences.

\begin{figure}[]
\epsscale{1.0}
\plotone{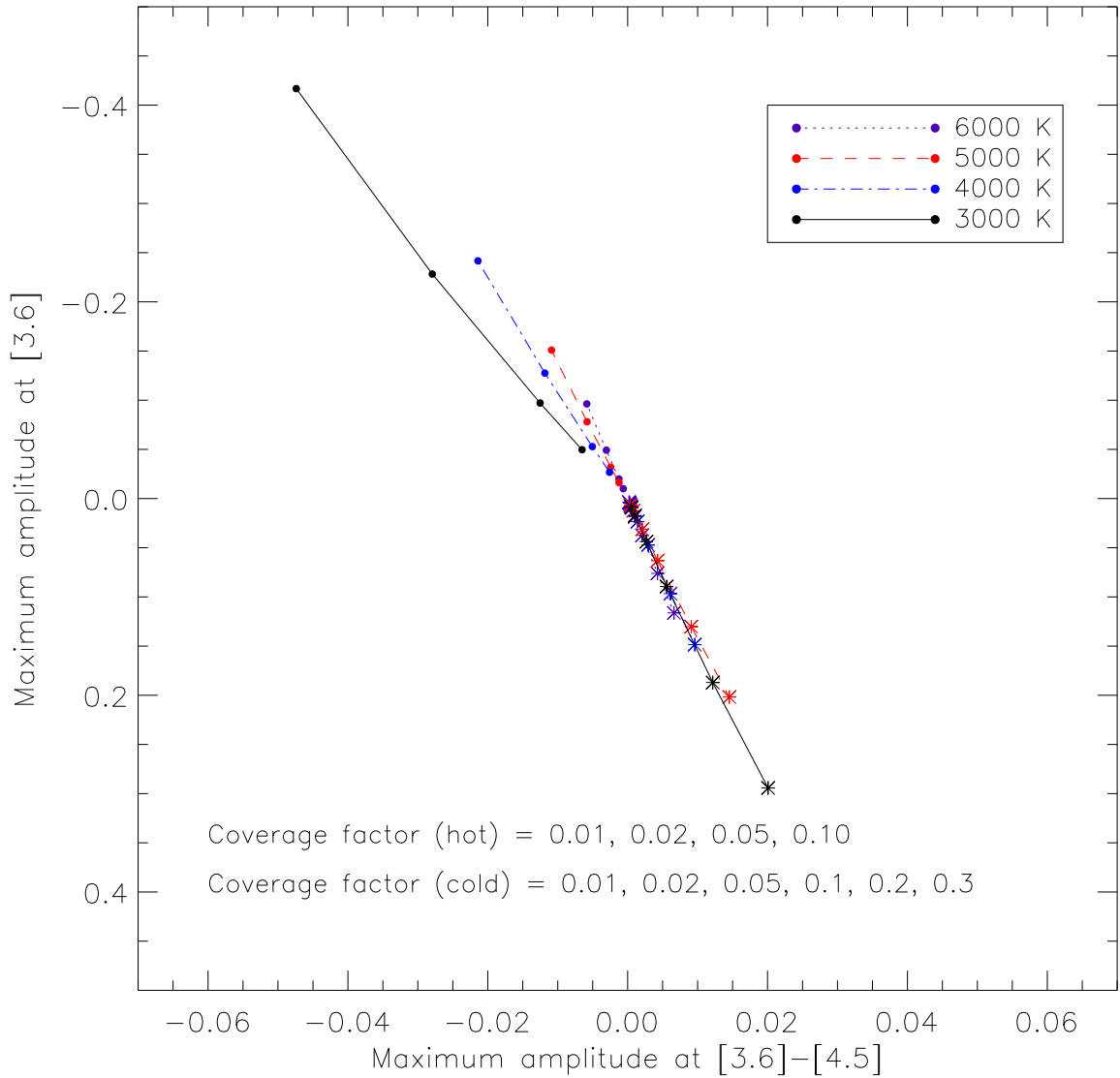}
\caption{Maximum amplitudes predicted by the spot model for a set of stars with temperatures ranging from 3000~K to 6000~K. Filled circles and asterisks represent the amplitudes predicted for hot and cold spots respectively.
}
\label{fig:spots}
\end{figure}

\begin{figure}[]
\epsscale{0.6}
\plotone{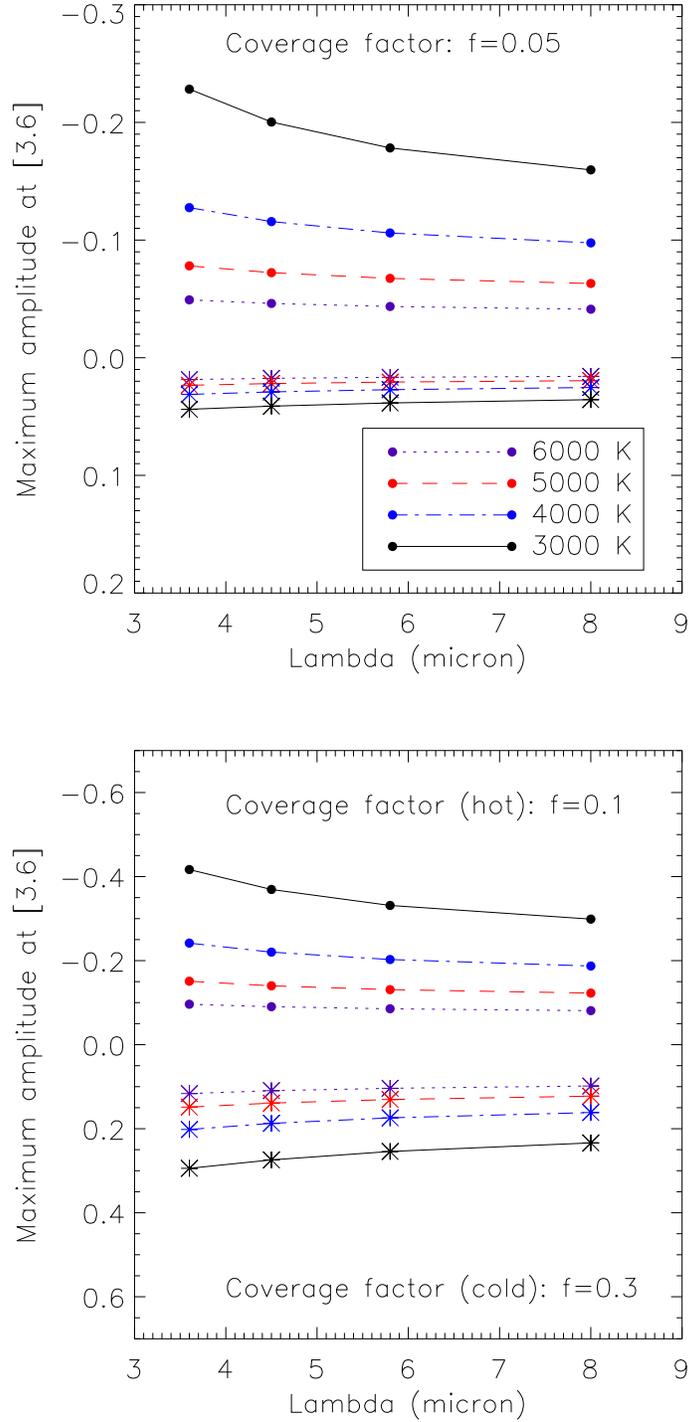}
\caption{Maximum amplitudes predicted by the spot model for a set of stars with temperatures ranging from 3000~K to 6000~K. Filled circles and asterisks represent the amplitudes predicted for hot and cold spots respectively. The upper panel show intermediate coverage factor while the bottom panel presents a more extreme situation with a coverage factor of 0.1 for hot spots and a coverage factor of 0.3 for cold spots.
}
\label{fig:spotsvslambda}
\end{figure}

Figure~\ref{fig:spots} shows the maximum predicted amplitudes at 3.6~$\mu$m 
and color [3.6]-[4.5] for typical stars with temperatures ranging between 
3000~K and 6000~K (expected temperatures for 2MASS J magnitudes between 
9.5 and 14~mag at 900~pc if one assumes an age of 0.5~Myr \citep{Siess00}). 
We assumed that hot spots (filled circles) have temperatures of order 10000~K, 
and cold spots (represented by asterisks) have temperatures 1500~K cooler than 
the photosphere. Different coverage fractions with a maximum of 10\% for 
hot spots, and up to 30\% for cold spots are shown. In Figure~\ref{fig:spotsvslambda} 
the maximum amplitude of variation for each channel is shown. These diagrams 
show that the difference in amplitude between different bandpasses is very 
small, as is true for our light curves, and the
amplitudes are smaller for higher photospheric temperatures.

However, these simple spot models are 
missing an important contribution relevant
to our case.  Essentially all of our objects are Class~I and
Class~II YSO's, with significant flux excesses over photospheric at IRAC
wavelengths.  The \citet{Vrba86} model assumes all of the flux is coming from
photosphere and spots.  If warm dust in the disk or envelope
contribute substantially to the integrated fluxes at 3.6~$\mu$m onward,
and if the disk/envelope flux is invariant with time, then the variability
amplitudes we observe with IRAC would be substantially lower than predicted
by equation {2}.

   We consider two extreme cases.  At one extreme, we consider the   case 
where nearly all of the light we see from the variable YSO's   is scattered 
light.  The apparent excess fluxes at IRAC wavelengths   are therefore 
instead indicative of lower extinction to the scattering   surface at longer 
wavelengths, rather than thermal emission from   dust in the inner disk/envelope.  
Since all of the light we see originates   from the star, the IRAC light 
curve amplitudes could be nearly as large as   the model estimates, dependent 
on the geometry assumed for the scattering   surface(s).  \citeauthor{Wood98} 
have constructed models of Class I YSOs with hot spots illuminating a 
disk+envelope, and have shown that even when the direct light from the star 
is heavily attenuated, the photometric signature of the hot spot (as it 
rotates with the stellar photosphere) can be seen in the scattered light 
flux from the envelope, at least at I band. For a model where the star+hot 
spot had an I band light curve amplitude of around 1 mag and where 
direct light from the star was almost completely blocked by the disk 
(i = 82 deg), the scattered light at I band showed a light curve amplitude 
of about 0.2 mag. A possible prototype for our IC1396A variable objects 
is HL Tau, where it is believed that essentially all the flux we see at 
least out to 2~$\mu$m is scattered light. If scattered light dominates 
out to 8~$\mu$m, then perhaps the \citeauthor{Wood98} 
model could provide the correct physical model to explain our light curves. 
With these assumptions, the observed flux at IRAC Ch. 4 provides a lower limit
   to the photospheric flux of the star at 8~$\mu$m.  For our stars with 
   quasi-sinusoidal light curves, and assuming an age of 0.5 Myr from Siess 2000,
   this suggests masses of order 0.3 to 4.5~M$_\sun$, and spectral types of M5 to F3 with the majority of the objects having masses around 3.5~M$_\sun$ and spectral types $\sim$K3.

However, it is not obvious that the variable YSOs in IC1396A are as evolutionarily 
young as HL Tau. Some of them fall in the IRAC color-color diagram within 
the Class II region, and do not necessarily have any significant remaining 
infalling envelope to serve as the scattering surface\footnote{We note, 
however, that according to the models of \citet{Robitaille07}, some Class I YSO's have colors
that fall in the Class II region of the IRAC CCD, and therefore it is possible that all of our
"periodic" variables may be Class I's.}. 
Also, it is difficult to understand how these stars could have such extreme
hot spots (much larger covering factor than typically required to explain
CTT light curves in the optical or near-IR) and hence presumably very 
active accretion without having a significant amount of warm dust 
contributing strongly to the flux at IRAC wavelengths.
Therefore, as the other extreme model, we consider a case where the observed 
J band magnitude is assumed to provide a reasonable estimate for the 
photospheric flux of the YSO (an assumption often used when attempting to 
place YSO's in HR diagrams to estimate their ages).  The fluxes at IRAC 
wavelengths are then the sum of a photospheric component (estimated by 
taking the J band flux and extrapolating to longer wavelengths with a 
black-body function) plus excess emission from warm dust.   The fraction 
of the IRAC flux that is emitted by the star then places constraints on 
models to explain the light curve amplitudes.  
As an example, we derive $F_{phot}$ = 0.09, 0.04, 0.02, and 0.01 for IRAC Ch.~1 
through Ch.~4 for object IC1396A-47 (See Fig.~\ref{fig:zeta}), where $F_{phot}$ 
is the fraction of the total system flux coming from the star in each of the IRAC bands. 
In the absence of any other mechanisms, and assuming the disk emission is 
invariant with time, if the star had an amplitude in its IRAC light curves 
of 1~mag, the star$+$disk would have amplitudes of only of order 0.09 mag 
in Ch1 and 0.01 mag in Ch4. Because no plausible spot model produces light 
curve amplitudes anywhere near 1~mag at IRAC wavelengths (for just the star 
alone), and because we see no wavelength dependent variation in amplitude 
for objects similar to  IC1396A-47, we conclude that this simplest star+spot 
plus disk model cannot account for our observed light curves. 

\subsection{Variable heating of the disk and disk inhomogeneities}\label{disk heating}
%
%

Time varying heating of the inner disk by the rotating spot could provide
 the mechanism for IRAC variability, if some property of the viewing geometry
 allows our view of the warmer part of the disk to vary with time.
 In this case, it is the luminosity of the spot compared to the luminosity 
of the rest of    the photosphere that matters, so if the temperature of 
the hot    spot is twice that of the photosphere, a spot covering fraction    
of only a little more than 1\% could yield a 0.2 mag light curve    amplitude from the disk. 
For this flavor of models, and assuming a medium Av of 8 magnitudes, the J band fluxes correspond to a mass range
of 0.2 to 1.3 solar mass for our periodic YSO's.

We have constructed  models of a Class~I protostar (with a disk and
envelope,
and a photosphere with a hot spot) using the radiative transfer code
described
in \citet{Whitney03a,Whitney03b}, and have varied the inputs to these models,
including the inclination angle at which we view the disk. If there are hot spots in the photosphere of the object, the disk will be asymmetrically heated too, and that could help keep the variations at IRAC bands at a higher level than if the disk emission were a constant source of additional flux.  Figure~\ref{fig:barbmodels} shows the results of these models for two viewing angles, with the temporal variation
of the spectral energy distribution in the left panels, and the magnitude variations
in specific bands for one full rotation period of the star in the right panels (asterisk:  V-band, diamond:  R-band, triangle: [3.6], box:  [4.5], open circle: [5.8], and filled circle: [8.0]). A star of mass 1.5 M$_\sun$ with effective temperature $\sim$4300K was used.  For accretion parameters, we used  a disk mass of 0.05~M$_\sun$, a disk accretion rate of  8.6$\cdot$10$^{-7}$M$_\sun$/yr, and an envelope accretion
rate of 4$\cdot$10$^{-6}$~M$_\sun$/yr. A hot spot with a fractional area coverage of 0.3\% and a temperature of 10000K  sitting at latitude 45 degrees was added to the models. The figure shows that these models can easily reproduce the IRAC light curves with the right amplitudes with
very modest hot spot sizes.  

 Close examination of the details of the model show that the
IRAC variations arise from the heating of the inner disk wall (that is, the IRAC
light curves are brightest when the photospheric hot spot is on the far side of
the star, heating up the back wall).  The IRAC variations are largest at close-to-edge-on
viewing angles where the far side of the wall is in view.  At more pole-on inclinations (bottom panels of
figure~\ref{fig:barbmodels}), the flux from the inner disk shows less phase dependence of our view of the hotter part of the inner disk so contribute less to the overall flux which consequently shows smaller
variations.  In the optical, the star is brightest when the hot spot is facing the observer.   Therefore, the
optical and IRAC light curves are 180 degrees out of phase. 

Similar results are obtained for a Class~II source as shown in Figure~\ref{fig:barbmodels2}.  The top-left panel
shows the same stellar, disk, and hotspot parameters as in Figure \ref{fig:barbmodels} but with no envelope.
The IRAC variations are similar (the y-axis scales are different), but the visible light curve shows much larger
variation in the Class~II source.  In the Class~I source, the visible light is scattered in the envelope, which decreases
and washes out the variations.  In the Class~II source, we clearly see the stellar hotspot rotate out of view and then stay at the same brightness until it rotates into view.

The top right panel of Figure \ref{fig:barbmodels2} adds a warp to the disk:  at the same longitude as the stellar
hotspot, the disk height increases by 25\% over the nominal value (we use a $\sin^{11}(\Lambda)$ function, where $\Lambda$ is the longitude, to make a smooth
transition).  
The warp decays exponentially with radius, with an e-folding distance equal to the inner disk radius $R_{in}$; thus, at a radius of several $R_{in}$, the disk becomes axisymmetric.  
The warp is meant to simulate the effect of an accretion flow to the stellar hotspot, and dragging some dust with it,
effectively increasing the height of the disk.
A model of this type has previously been proposed to
explain the observed optical variability of AA Tau \citep{Bouvier97,Bouvier03}. 

The result of this is that the IRAC variations are larger except at 8 $\mu$m.  
This is because the regions of the wall that is heated by the stellar hotspot has a bigger area than the model with no warp.
The optical variations are lower with the inclusion of the warp.   
This is because the reflection off the warped wall adds flux back into the beam as the hotspot moves out of view.

The bottom panels of Figure \ref{fig:barbmodels2} show two stellar hotspots separated in longitude by 180\arcdeg,
with one at +45\arcdeg latitude and the other at -45\arcdeg, and disk warps at these longitudes (one in the $+z$ direction and the
other in the $-z$).  These are shown for two different viewing angles (bottom left and right panels).
The IRAC light curves are now double peaked due to the two warps.  
The IRAC variations are similar in the two viewing angles shown (bottom left and right panels), while the visible
light curves show large differences.  The optical light curves are more complicated due to the geometry of the stellar hotspots, and scattering and obscuration from the warped wall. 

These models demonstrate that both hotspots and disk inhomogeneities can explain the 
variability in the 30\% of our sample of YSOs that shows 5-12 day periodic light curves. 
The amplitudes and behaviors are a function of the spot temperature and size, number of hotspots,
viewing angle, and possibly inner disk radius (the latter has not been investigated yet).
Multiwavelength observations combined with models should be able to distinguish between single- and double-hotspot
models and viewing angles.

\subsection{IC1396A-47: Pulsating pre-main sequence star?} \label{zeta}
%
%

Our object IC1396A-47 was previously identified in \citet{Reach04} as 
object $\zeta$. It has been classified by us and others as a Class~I YSO based on
its IRAC SED shape and position in the IRAC CCD. However, when we
fit its SED using an evolutionary sequence of models from Class 0 to Class III \citep{Whitney03a,Whitney03b}, object 47 is best fit with a pure
disk model seen nearly edge on, with essentially no envelope, and with a large extinction
to explain the very red J-K color. The primary reason why an edge-on Class II model is favored is the low 24~$\mu$m flux.

\begin{figure*}[]
\centering
\includegraphics[angle=0,scale=0.30]{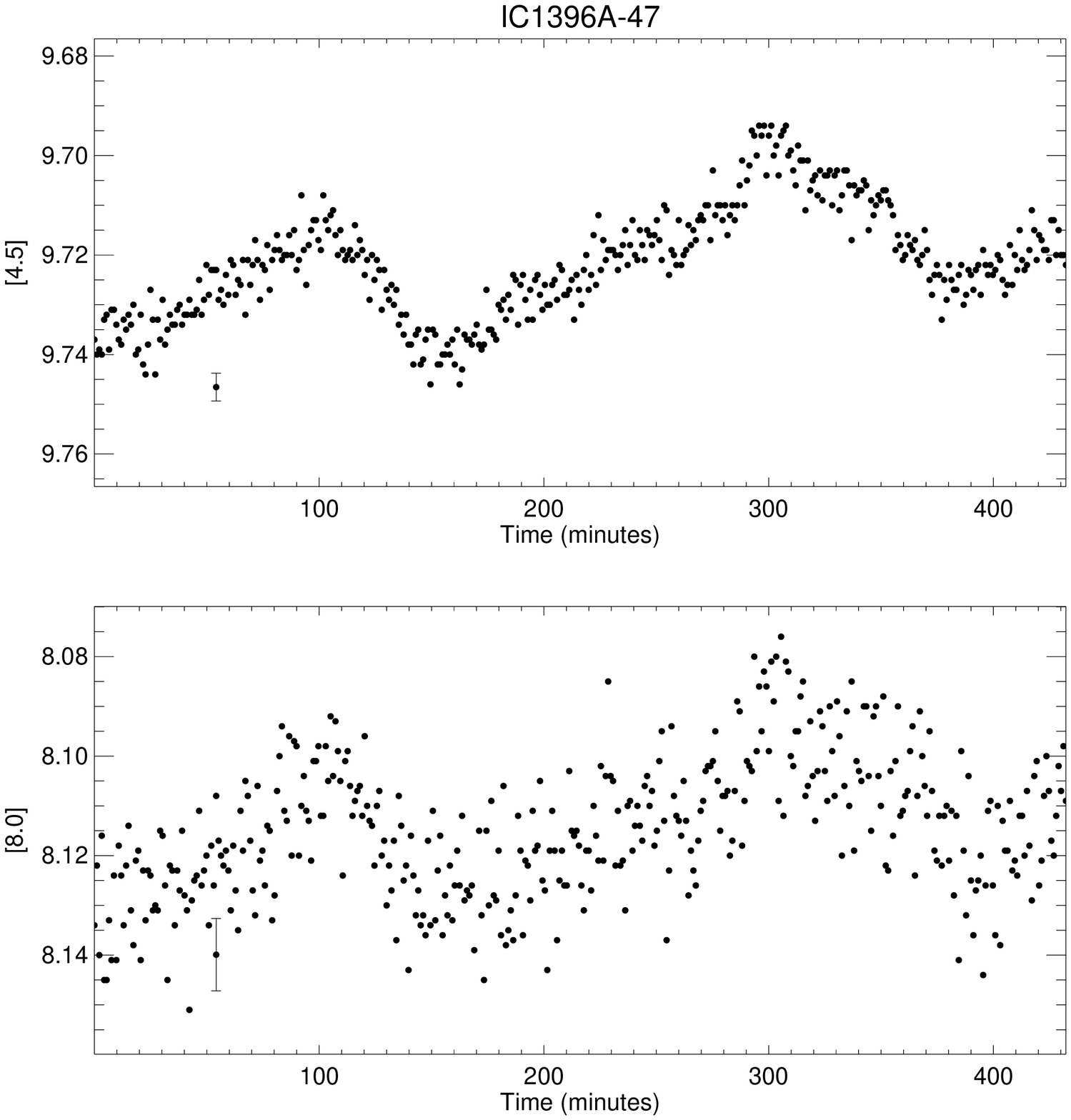}
\includegraphics[angle=0,scale=0.50]{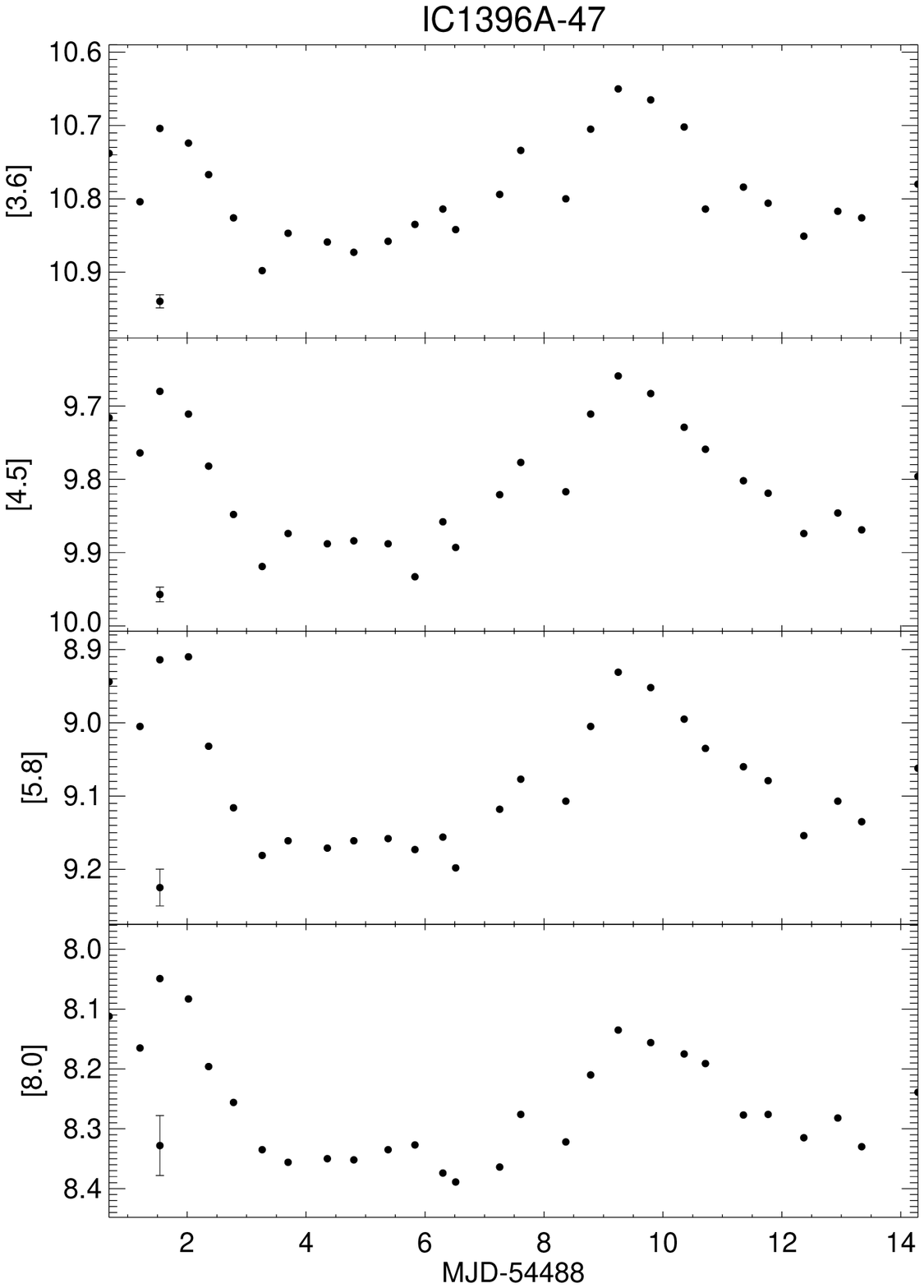}
\caption{{\bf a)}Light curves of IC1396A-47 (object $\zeta$) showing  the 3.5 hours period variation from the 7 hour staring-mode IRAC
observation. The top and bottom panels show Ch. 2 and Ch. 4 time series,
respectively. {\bf b)} Light curves of the same object from IRAC mapping observations showing the $\sim$9 days period variation in  the four IRAC bandpasses.}
\label{fig:zeta}
\end{figure*}

IC1396A-47 shows a periodic variation with a peak-to-peak amplitude 
of 0.2 mag and a derived period of 9 days in the mapping data (See Fig.~\ref{fig:zeta}b). 
However, this object was the only one also showing a periodic short term 
variation. As shown in  Fig~\ref{fig:zeta}a, it exhibits an apparent 
$\sim$ 3.5~hr periodicity with amplitude of order 0.04 mag.  
Contact binaries can have similar periods and light curve shapes; however, 
for the mass and radius required for object zeta, the apparent period is 
too short.  We believe that the most plausible explanation for the short 
period variability is that object zeta is a PMS delta Scuti (pulsational) variable.
A few PMS $\delta$ Scuti pulsators have been identified and 
some new possible members have also been proposed \citep{Rodriguez01,Zwintz05, Zwintz08b}. 
\citet{Breger72} identified two PMS $\delta$ Scuti variables in 
NGC2264 -- both have colors of F stars, periods of order 3 hours and V band 
light curve amplitudes of order 0.05 mag. 

If the J band flux of IC1396A-47 were photospheric, an A(V) of 50 mag would be required to allow zeta
to have the minimum mass to fall within the instability strip at 0.5 Myr \citep{Marconi98}.
If the Ch.4 flux is purely photospheric (see Sec.~\ref{variabilitycauses}), the derived mass for this object is $\sim$4-4.5M$_\sun$ (spectral type F3 and implied Av=35) which is compatible with it being in the instability strip.

\section{SUMMARY AND CONCLUSIONS} 
%
%

We have conducted a variability study using Spitzer/IRAC data keeping in mind that mid-IR variability probes different physical mechanisms than variability in the optical or near-IR. Because IRAC is sensitive to the dust in the inner disk and envelope, variability at IRAC wavelengths in YSOs should primarily trace changes in the heating and  fluctuations in the structure of the circumstellar disk and envelope surrounding the nascent star whereas optical and near-IR variability is directly related to the central object.  Detailed IRAC light curves thus have the potential to allow us to construct models of the structure of the inner disk and the temporal variations in disk structure and mass accretion. Some of these variations could be tied to either the formation or migration of planets 
\citep{Ida08}. Therefore, mid-IR variability offers a new means to infer hows stars are assembled and
the possible early stages of planet formation.

Our photometric monitoring program on the IC1396A dark 
globule focuses on studying the mid-IR variability of the heavily 
embedded YSOs present in that area. Observations were performed with 
Spitzer/IRAC using two different setups corresponding to distinct timespan and temporal cadences: i) light curves covering a 14 day timespan with a twice daily
cadence for 69 YSOs, and ii) continuous light curves with approximately 12 second cadence over 7 hours for 38 YSOs. Typical accuracies for our 
relative photometry were of order 3~mmag for the short term data, and around 10~mmag for the longer term data. This is the only star-forming region for which such extensive time series 
photometry has been obtained with Spitzer. 

\begin{figure*}[]
\centering
\includegraphics[angle=0,scale=1.0]{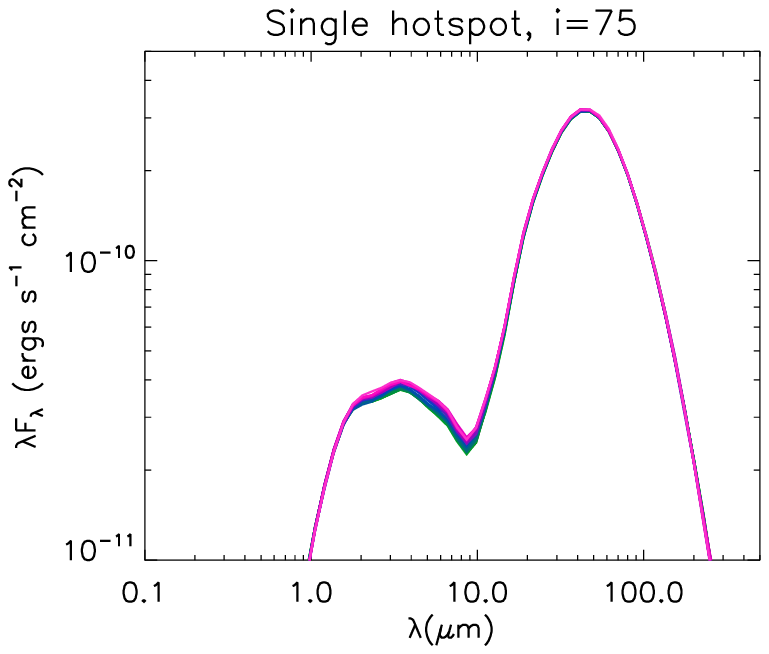}
\includegraphics[angle=0,scale=1.0]{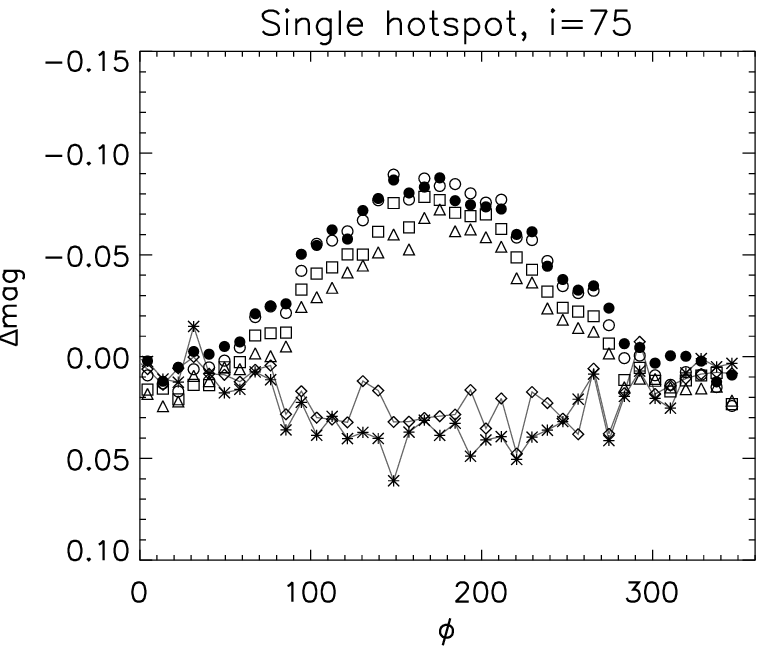}
\includegraphics[angle=0,scale=1.0]{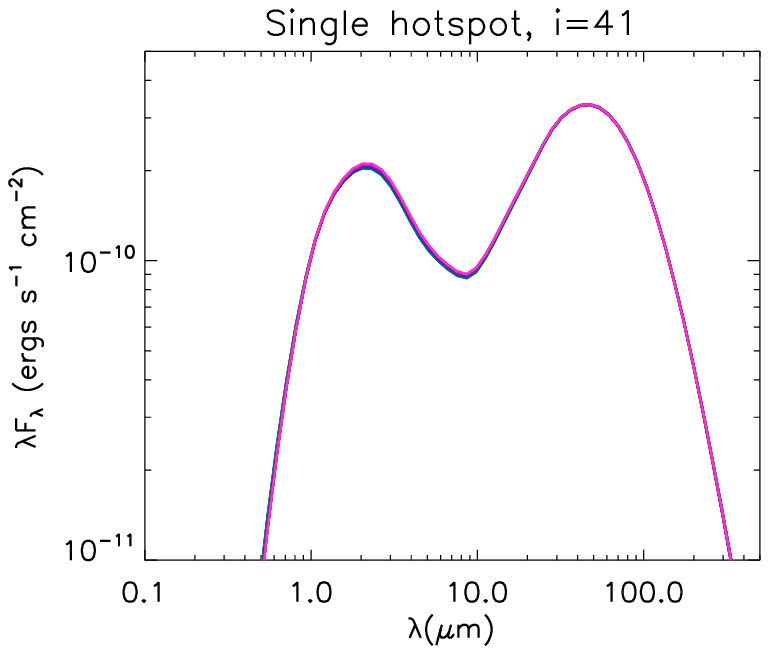}
\includegraphics[angle=0,scale=1.0]{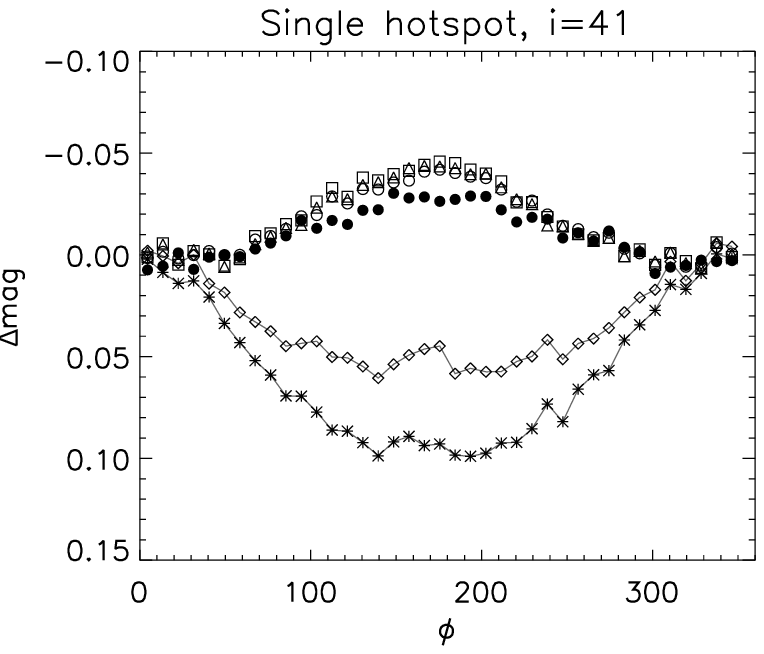}
\caption{Model results for a Class~I object with a hot spot on the photosphere, at a latitude of $45\arcdeg$. The different
color SEDs (left) correspond to different phase angles; and the light curves (right) show one full rotation period at various wavebands (asterisk: V-band, diamond: R-band, triangle: [3.6], box: [4.5], open circle: [5.8], and filled circle: [8.0]). 
The V- and R-band lightcurves are connected by lines to distinguish them from the IR lightcurves. Phase 0 corresponds to the stellar hotspot facing the observer.  At phase=180 degrees, the hotspot is obscured and the heated far wall faces the observer.  Positive $\Delta$mag corresponds to fainter fluxes.
The visible light curves are noisier and washed out due to extinction and scattering from the envelope.
The result are periodic colorless
IR variations with amplitudes similar to those that we observe at IRAC bands, and  with the brightest optical phase corresponding to the faintest IRAC phase.  
At more pole-on viewing angles (bottom panels), the visible light curves show more variation, and the IR lightcurves show less.  
This is because the stellar hotspot, located at 45 degrees latitude, shows more pronounced variation with rotation, whereas the flux from the inner disk shows less phase dependence with more pole-on viewing angle.
}
\label{fig:barbmodels}
\end{figure*}

\begin{figure*}[]
\centering
\includegraphics[angle=0,scale=1.0]{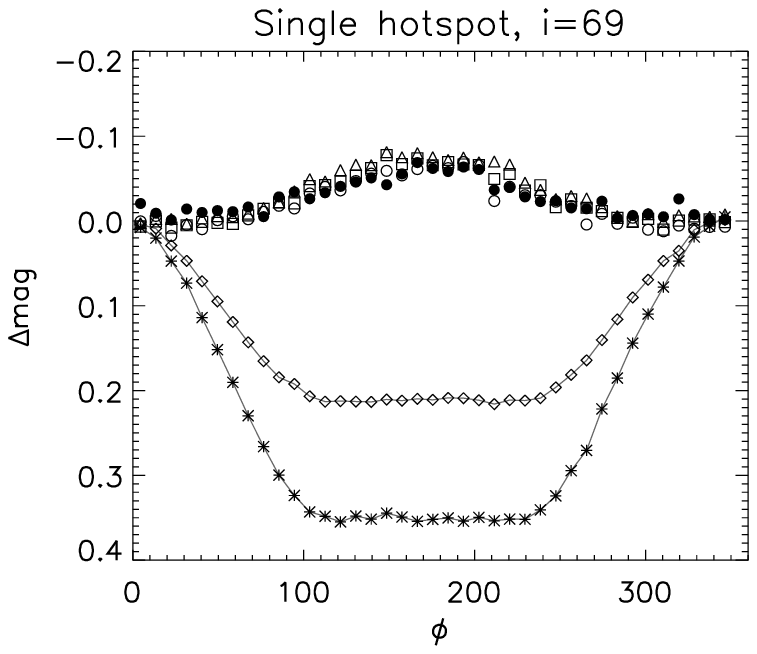}
\includegraphics[angle=0,scale=1.0]{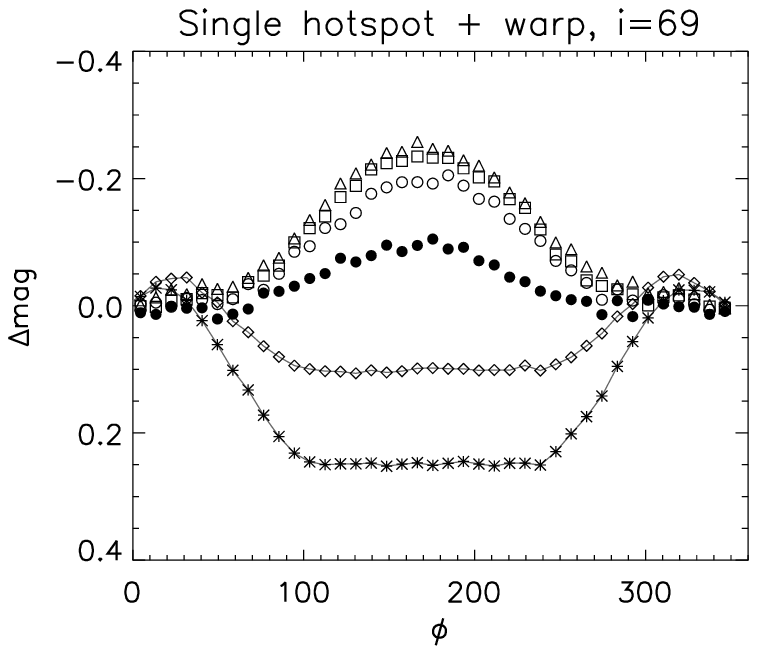}
\includegraphics[angle=0,scale=1.0]{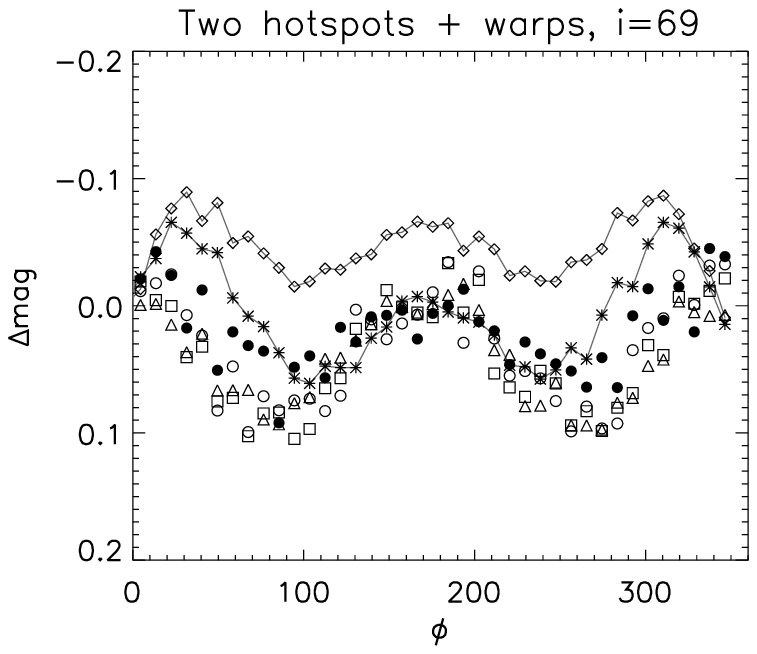}
\includegraphics[angle=0,scale=1.0]{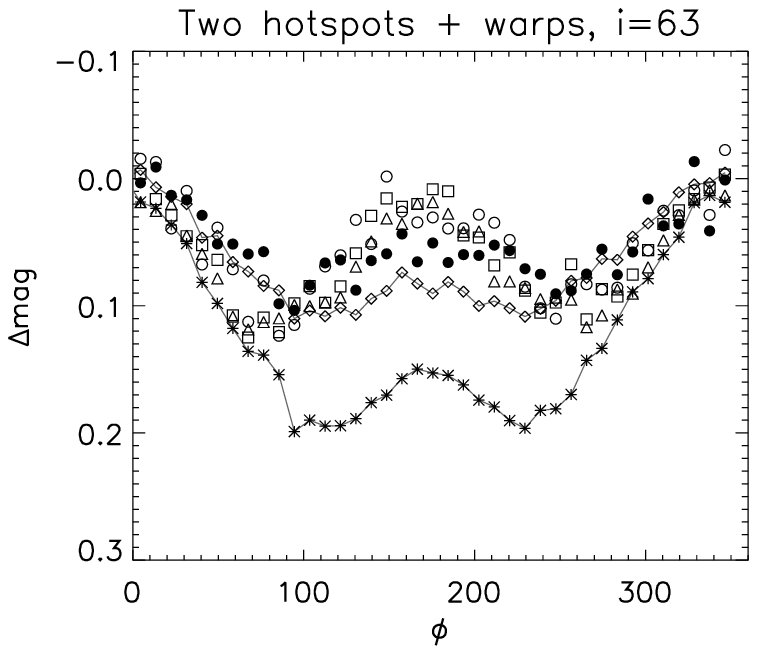}
\caption{Hot spot model results for a Class II source.  The top left model has the same stellar, disk, and hotspot parameters as Figure \ref{fig:barbmodels}, but no envelope.    The visible light curves show more variation without the envelope.  
The top right model adds a warp in the disk, raising the disk height 25\% at the longitude of the hotspot.
This decreases the visible variation and increases the IR.
The bottom two models have 2 spots located 180\arcdeg apart in longitude and disk warps at the spot longitudes.
The visible light curves vary more with inclination (bottom left
and bottom right).
The symbols are as in previous Figure.  
}
\label{fig:barbmodels2}
\end{figure*}

Around 60\% of our sample of YSOs (69 objects) were determined to be
variable on long timescales 
and 8 objects (out of the 38 objects with continuous light curves over a 7~hr timespan) are also labeled as variable in short timescales.  We have 
found very different shapes in the long term light curves but the most noticeable 
characteristic is the very good similarity among the light curves in 
different bandpasses both in shape and in the amplitude of the variation. 
About 30\% of our total sample of YSOs show 
periodic-like, colorless variability with amplitudes up to 
about 0.2 mag. These quasi-periodic objects are about equally divided between Class~I and Class~II and have possible masses ranging from 0.2 to $\sim$4.0~M$_\sun$ (depending on
the assumed extinction). We have investigated different scenarios to understand the 
cause of this type of variability, variable accretion, hot and cold spots 
in the photosphere of the objects, and variable obscuration produced by 
the circumstellar disk. We believe that the most plausible cause of variability for these objects are hot spots on the photosphere of our targets which asymmetrically heat the inner wall of the disk (Fig. \ref{fig:barbmodels}).  Disk warping can enhance this effect (Fig. \ref{fig:barbmodels2}). We have used radiation transfer models to check this scenario and the models are able to reproduce the observed variability in the four IRAC bands with only small changes in accretion and/or inclination angle needed to get the range of amplitudes found in our YSOs. 

One of our targets, IC1396A-47, was the only one also showing 
periodic short term variation with an apparent $\sim$~3.5 hr periodicity 
and an amplitude of order 0.04~mag.  We propose that IC1396A-47 is a PMS 
delta Scuti variable. A dedicated, long duration photometric monitoring campaign on this
object could in principle determine its mass and age from its pulsation
frequencies, thereby testing theoretical PMS models at very young ages.

In addition to the variability 
study we have searched for new IC1396A candidate members using two different methodologies. On the one hand, we co-added the observations to obtain a deeper mosaic and 
extracted 15 new candidate members, 13 out of them fall into the Class~II 
region in the IRAC color-color diagram and 2 are classified as Class~I 
objects.  On the other hand, we  have searched for variability, as an indicator of 
youth, in the field and have detected around 50 possible new Class~III candidates.  

These observations have shown that Spitzer/IRAC
is an exquisite instrument to carry out very accurate photometric monitorings both in timescales of hours and days. During the post-cryogenic operations, it will be the prime means to carry out this type of studies until the arrival of the James Webb Space Telescope.



\acknowledgments
This work is based [in part] on observations made with the Spitzer Space Telescope, which is operated by the Jet Propulsion Laboratory, California Institute of Technology under a contract with NASA. This research has been funded by the NASA grant JPL101185-07.E.7991.020.6, and the Spanish grants MEC/ESP 2007-65475-C02-02, MEC/Consolider-CSD2006-0070, and CAM/PRICIT-S-0505/ESP/0361. MMC acknowledges the support by a predoctoral "Calvo Rode\'es" fellowship by INTA.



{\it Facilities:} \facility{Spitzer (IRAC)}





\clearpage
%

\clearpage

\begin{deluxetable}{lcccccccccc}
\tabletypesize{\tiny}
\rotate
\tablecolumns{18}
\tablewidth{0pt}
\tablecaption{New candidates photometry from the co-added deep mosaics.\label{tab:newphot}}
\tablehead{ 
\colhead{Object\tablenotemark{a}} &\colhead{RA(J2000), DEC(J2000)} &\colhead{J\tablenotemark{b} eJ} &\colhead{H\tablenotemark{b} eH} &\colhead{$K_s$\tablenotemark{b} e$K_s$}&\colhead{[3.6] e[3.6]} &\colhead{[4.5] e[4.5]} &\colhead{[5.8] e[5.8]} &\colhead{[8.0] e[8.0]}&\colhead{[24] e[24]} &\colhead{Class}}
\startdata
IC1396A-60  & 21:36:47.18  +57:29:52.6 & 14.184\tablenotemark{c}  --     & 14.073  0.058  & 12.63   0.028  & 10.419  0.003 & 9.874   0.003 & 9.419   0.008 & 8.846   0.02  & -- -- &II   \\
IC1396A-61  & 21:36:47.63  +57:29:54.1 & 13.568\tablenotemark{c}  --     & 12.342  0.042  & 11.655  0.033  & 10.426  0.004 & 9.964   0.004 & 9.545   0.01  & 8.972   0.031 & -- -- & II   \\
IC1396A-62  & 21:37:14.51  +57:28:40.9 & 14.511  0.039  & 13.559  0.041  & 12.954  0.034  & 11.839  0.002 & 11.349  0.002 & 11.128  0.002 & 10.609  0.01  & 7.012 0.039 & II   \\
IC1396A-63  & 21:37:17.42  +57:29:27.5 & 14.122  0.043  & 13.135  0.043  & 12.583  0.037  & 11.859  0.003 & 11.436  0.004 & 11.104  0.003 & 10.249  0.008 & 7.165 0.054 & II   \\
IC1396A-64\tablenotemark{d}  & 21:37:17.37  +57:29:20.8 & 14.091\tablenotemark{c}  --     & 13.557  0.071  & 13.13\tablenotemark{c}   --     & 12.901  0.012 & 12.669  0.013 & 12.344  0.011 & 11.528  0.023 & -- -- & II   \\
IC1396A-65  & 21:36:42.49  +57:25:23.3 & 14.806  0.043  & 14.089  0.047  & 13.758  0.049  & 13.194  0.007 & 12.956  0.007 & 12.553  0.01  & 11.928  0.041 & 8.959 0.158 & II   \\
IC1396A-66  & 21:37:05.87  +57:32:12.5 & 15.13   0.051  & 14.284  0.059  & 14.139  0.078  & 13.372  0.012 & 13.086  0.012 & 12.8    0.066 & 11.953  0.123& -- -- & II   \\
IC1396A-67  & 21:35:58.52  +57:29:15.1 & 16.582  0.167  & 15.373  0.125  & 14.632  0.1    & 13.537  0.019 & 13.088  0.013 & 12.534  0.04  & 11.556  0.121& -- -- & II   \\
IC1396A-68  & 21:36:56.26  +57:29:52.3 & 18.527\tablenotemark{c}  --     & 16.098\tablenotemark{c}  --     & 15.2    0.141  & 13.577  0.03  & 12.981  0.022 & 12.15   0.096 & 11.126  0.224& -- -- & II   \\
IC1396A-69  & 21:36:38.03  +57:26:57.9 & 15.604  0.072  & 14.893  0.073  & 14.523  0.08   & 13.58   0.009 & 13.158  0.007 & 12.796  0.011 & 11.979  0.02  &8.179 0.137 & II   \\
IC1396A-70  & 21:36:12.60  +57:31:26.3 & 16.504  0.132  & 16.161\tablenotemark{c}  --     & 15.136  0.14   & 14.006  0.019 & 13.481  0.012 & 13.34   0.077 & 12.429  0.195 & -- -- &II   \\
IC1396A-71  & 21:36:40.34  +57:25:45.7 & 16.271  0.094  & 15.203  0.099  & 14.867  0.113  & 14.059  0.009 & 13.618  0.008 & 13.203  0.016 & 12.366  0.022 &9.051 0.149 & II   \\
IC1396A-72  & 21:36:18.97  +57:29:05.1 & --      --     & --      --     & --      --     & 14.197  0.014 & 13.157  0.007 & 12.29   0.022 & 11.297  0.059 & -- -- &I    \\
IC1396A-73  & 21:37:11.78  +57:30:34.9 & 16.175  0.115  & 15.52\tablenotemark{c}   --     & 14.928  0.148  & 14.289  0.013 & 13.823  0.015 & 13.372  0.036 & 12.462  0.057 &6.305 0.072 & II   \\
IC1396A-74  & 21:36:36.35  +57:32:09.3 & --      --     & --      --     & --      --     & 15.505  0.085 & 14.679  0.02  & 13.237  0.104 & 11.606  0.101 & -- -- &I    \\
\enddata
\tablenotetext{a}{\cite{Sicilia06} identified 57 YSOs in this region and there are another 2 objects that were identified by \citet{Reach05} thus we name our new candidate members begining with number 60.}
\tablenotetext{b}{Photometry for the J, H and $K_s$ bandpasses has been taken from the 2MASS database \citep{Cutri03}.}
\tablenotetext{c}{Upper limmits}
\tablenotetext{d}{This object is very faint at 24~$\mu$m and it is located too close to a brighter object to derive a reliable magnitude.}
\end{deluxetable}


\clearpage

\begin{deluxetable}{lcccccccccccccc}
\tabletypesize{\tiny}
\rotate
\tablecolumns{15}
\tablewidth{0pt}
\tablecaption{Long-term variability: Main results from mapping data.\label{tab:longvar}}
\tablehead{
\colhead{} & \colhead{} & \colhead{} & \colhead{} & \colhead{} &\multicolumn{4}{c}{3.6~$\mu$m} &  \multicolumn{4}{c}{4.5~$\mu$m} & \colhead{} & \colhead{}  \\
\colhead{Object\tablenotemark{a}} & \colhead{RA DEC} & \colhead{Class\tablenotemark{b}} & \colhead{ID,Class\tablenotemark{c}} & \colhead{Class\tablenotemark{d}} &\multicolumn{4}{c}{---------------------------------------------------} &  \multicolumn{4}{c}{---------------------------------------------------} & \colhead{Jindex} & \colhead{Var}\\
\colhead{} &\colhead{hr, deg (J2000)} & \colhead{SA06 } & \colhead{R04 } & \colhead{} &\colhead{mag} &\colhead{RMS} &\colhead{$\chi^2$} &\colhead{$ T_{var}$ (days)} &\colhead{mag} &\colhead{RMS} &\colhead{$\chi^2$}  &\colhead{$ T_{var}$ (days)} &\colhead{} &\colhead{Final}}
\startdata
IC1396A-1 &21:35:57.93 +57:29:09.9 &II      &$\xi$,I/II       &I\tablenotemark{f}  & 13.378$\pm$0.015 & 0.101 & V &- & 12.752$\pm$0.017 & 0.106 & V   & -         &2.46&          V     \\
IC1396A-2 &21:35:59.05 +57:30:23.3 &II      &           &II & 12.516$\pm$0.012 & 0.023 & V &- & 12.200$\pm$0.013 & 0.026 & V   &-                                 &0.33&          V     \\
IC1396A-3 &21:36:06.06 +57:26:34.1 &II      &           & I\tablenotemark{f} & 13.010$\pm$0.015 & 0.02  & N &- & 12.567$\pm$0.016 & 0.024 & N   &-                &-0.04&          N     \\
IC1396A-4 &21:36:07.98 +57:26:37.1 &I       &$\gamma$,I/0   & I & 10.805$\pm$0.006 & 0.061 & V &- &  9.026$\pm$0.009 & 0.048 & V   &-                             &2.20&          V     \\
IC1396A-5 &21:36:03.89 +57:27:12.1 &II      &           & II& 11.710$\pm$0.009 & 0.009 & N &- & 11.652$\pm$0.009 & 0.014 & N   &-                                 &0.02&          N     \\
IC1396A-6 &21:36:07.46 +57:26:43.6 &I       &           & I& 12.836$\pm$0.013 & 0.032 & V & 4.9 &   11.027$\pm$0.009 & 0.022 &  V   &5.1                          &1.01&          V     \\
IC1396A-7 &21:36:18.36 +57:28:31.6 &I       &$\epsilon$,I/0   &I & 12.744$\pm$0.012 & 0.079 & V &- & 10.701$\pm$0.009 & 0.083 & V   &-                            &3.22&          V     \\
IC1396A-8 &21:36:19.42 +57:28:38.5 &I       &$\delta$,I/0     &I & 12.792$\pm$0.012 & 0.033 & V &- & 11.050$\pm$0.009 & 0.023 & V   &-                            &0.39&          V     \\
IC1396A-9\tablenotemark{e} &21:36:14.20 +57:27:37.9 &II   &           & III\tablenotemark{f} &  8.972$\pm$0.006 & 0.027 & N &- &  8.893$\pm$0.009 & 0.009 & N   &-&-0.03&          N     \\
IC1396A-10\tablenotemark{e} &21:36:14.20 +57:27:57.7 &II   &           &II &13.531$\pm$0.016 & 0.034 & N & - & 13.610$\pm$0.028 &0.036 & N & -                    &0.07&          N     \\
IC1396A-11 &21:36:16.64 +57:28:40.4 &I &           &I & 12.502$\pm$0.012 & 0.025 & V &- & 11.751$\pm$0.009 & 0.022 & V   &-                                       &0.31&          V     \\
IC1396A-12 &21:36:16.99 +57:26:39.9 &II     &           & II & 12.852$\pm$0.012 & 0.02  & N &- & 12.501$\pm$0.015 & 0.012 & N   &-                                &0.01&          N     \\
IC1396A-13 &21:36:23.68 +57:32:45.2 &II     &           &II & 12.589$\pm$0.012 & 0.075 & V &9.1 & 12.374$\pm$0.014 & 0.071 & V   &9.1                             &0.88&          V     \\
IC1396A-14 &21:36:25.07 +57:27:50.2 &II     &           & II& 12.545$\pm$0.012 & 0.039 & V &- & 12.041$\pm$0.012 & 0.05  & V   & -                                &1.14&          V     \\
IC1396A-15\tablenotemark{e} &21:36:33.00 +57:28:49.3 &II  &           &II & 13.765$\pm$0.018 & 0.029 & V &-  &  13.355$\pm$0.024 & 0.044 &  V   &-                &0.36&          V     \\
IC1396A-16 &21:36:35.31 +57:29:31.1 &II     &           &II & 11.356$\pm$0.006 & 0.038 & V &- & 10.993$\pm$0.009 & 0.042 & V   & -                                &1.61&          V     \\
IC1396A-17 &21:36:36.91 +57:31:32.6 &II     &$\iota$,II    &II & 11.102$\pm$0.006 & 0.033 & V &- & 10.625$\pm$0.009 & 0.035 & V   & -                             &0.81&          V     \\
IC1396A-18 &21:36:38.41 +57:29:17.4 &II     &           &II & 11.353$\pm$0.006 & 0.049 & V &- & 10.975$\pm$0.009 & 0.039 & V   & -                                &0.97&          V     \\
IC1396A-19 &21:36:39.15 +57:29:53.3 &II     &$\theta$,II    &II &  8.581$\pm$0.006 & 0.071 & V &- &  7.859$\pm$0.009 & 0.064 & V   & -                            &0.06&          N     \\
IC1396A-20 &21:36:41.46 +57:30:27.8 &II     &           &II & 12.773$\pm$0.012 & 0.017 & N &- & 12.383$\pm$0.014 & 0.014 & N   & -                                &0.12&          N     \\
IC1396A-21 &21:36:41.65 +57:32:17.5 &II     &           &I\tablenotemark{f} & 12.653$\pm$0.012 & 0.03  & V &- & 12.165$\pm$0.012 & 0.032 & V   &-                 &0.53&          V     \\
IC1396A-22 &21:36:43.98 +57:29:28.7 &II     &           &II & 11.301$\pm$0.006 & 0.069 & V &8.2 & 10.761$\pm$0.009 & 0.047 & V   &9                               &1.71&          V     \\
IC1396A-23\tablenotemark{e} &21:36:44.00 +57:28:46.8 &II  &           &II & 12.416$\pm$0.009 & 0.016 & N &- & 12.420$\pm$0.014 & 0.015 & N   & -                  &0.08&          N     \\
IC1396A-24 &21:36:45.96 +57:29:33.9 &I      &           &I  &  9.644$\pm$0.006 & 0.056 & V &6 &  9.114$\pm$0.009 & 0.064 & V   &6                                 &2.88&          V     \\
IC1396A-25 &21:36:46.60 +57:29:38.4 &I      &$\alpha$,I     &I & 10.217$\pm$0.006 & 0.011 & N &- &  8.776$\pm$0.009 & 0.012 & N   &-                              &0.42&          N     \\
IC1396A-28\tablenotemark{e} &21:36:47.88 +57:31:30.6 &I  &           &I & 12.137$\pm$0.009 & 0.059 & V &5.4 & 11.862$\pm$0.009 & 0.043 & V   &-                   &0.67&          V     \\
IC1396A-29 &21:36:49.41 +57:31:22.0 &II     &           &II &  9.446$\pm$0.006 & 0.04  & V &7.8 &  9.031$\pm$0.009 & 0.034 & V   & -                              &1.06&          V     \\
IC1396A-30 &21:36:50.72 +57:31:10.6 &II     &LkHa349a,III  &III\tablenotemark{f} &  8.688$\pm$0.006 & 0.033 & V &- &  8.439$\pm$0.009 & 0.035 & V   & -           &2.3E-4&        N     \\
IC1396A-31 &21:36:52.81 +57:29:43.8 &II     &           &II & 12.572$\pm$0.012 & 0.011 & N &- & 12.269$\pm$0.013 & 0.011 & N   &-                                 &0.08&          N     \\
IC1396A-32\tablenotemark{e} &21:36:54.50 +57:30:05.1 &II  &           &III\tablenotemark{f} & 10.108$\pm$0.006 & 0.019 & N &- &  9.889$\pm$0.009 & 0.006 & N  & - &-0.01&          N     \\
IC1396A-33\tablenotemark{e} &21:36:54.75 +57:31:45.0 &II  &           &II & 13.354$\pm$0.015 & 0.03  & N &- &    -- & --   & -   & -                              &0.40&          N     \\
IC1396A-34 &21:36:54.89 +57:30:00.3 &I      &$\lambda$,I   &I & 11.932$\pm$0.009 & 0.042 & V &- & 11.251$\pm$0.009 & 0.037 & V   &-                               &0.89&          V     \\
IC1396A-35 &21:36:55.20 +57:30:30.1 &II     &$\beta$,II   &II &  9.811$\pm$0.006 & 0.053 & V &- &  9.309$\pm$0.009 & 0.033 & V   & -                              &0.57&          V     \\
IC1396A-36 &21:36:55.43 +57:31:39.1 &II     &           &III\tablenotemark{f} & 11.222$\pm$0.006 & 0.009 & N &- & 11.187$\pm$0.009 & 0.008 & N   & -              &0.04&          N     \\
IC1396A-37 &21:36:56.98 +57:29:22.7 &II     &           &II & 11.569$\pm$0.009 & 0.049 & V &9.1 & 11.049$\pm$0.009 & 0.041 & V   &9.1                             &1.18&          V     \\
IC1396A-38 &21:36:57.67 +57:27:33.1 &II     &           & II&  9.447$\pm$0.006 & 0.033 & V &- &  8.987$\pm$0.009 & 0.049 & V   & -                                &1.12&          V     \\
IC1396A-39 &21:36:57.83 +57:30:56.1 &    I  &$\eta$,I      &I & 12.186$\pm$0.009 & 0.057 & V &7.8 & 11.139$\pm$0.009 & 0.058 & V   &9                             &2.26&          V     \\
IC1396A-40 &21:36:57.93 +57:29:10.7 &I      &$\nu$,I/II    &II\tablenotemark{f} & 10.316$\pm$0.006 & 0.025 & V &6 &  9.916$\pm$0.009 & 0.027 & V   &7.6           &0.60&          V     \\
IC1396A-41 &21:36:58.90 +57:30:29.2 &I      &           &I & 12.865$\pm$0.012 & 0.032 & V &- & 12.211$\pm$0.013 & 0.027 & V   & -                                 &0.19&          V     \\
IC1396A-43 &21:36:59.47 +57:31:34.9 &II     &           &II & 11.739$\pm$0.009 & 0.055 & V &10.5 & 11.251$\pm$0.009 & 0.063 & V   &-                              &1.90&          V     \\
IC1396A-44\tablenotemark{e} &21:37:01.05 +57:30:39.7&I  & & I & 13.257$\pm$0.015 & 0.023 &N&-&  13.110$\pm$0.021 & 0.026& N& -                                    &-0.06&          N     \\
IC1396A-45 &21:37:01.91 +57:28:22.2 &II     &           &II & 10.856$\pm$0.006 & 0.027 & V &6 & 10.535$\pm$0.009 & 0.034 & V   & 6                                &0.28&          V     \\
IC1396A-46 &21:37:02.00 +57:31:55.3 &II     &           &II & 12.061$\pm$0.009 & 0.017 & N &- & 11.911$\pm$0.009 & 0.015 & N   & -                                &0.13&          N     \\
IC1396A-47 &21:37:02.32 +57:31:15.2 &I      &$\zeta$,I   &I & 10.790$\pm$0.006 & 0.065 & V &9 &  9.807$\pm$0.009 & 0.078 & V   & 9                                &3.86&          V     \\
IC1396A-49 &21:37:06.49 +57:32:31.6 &II     &           &II & 10.900$\pm$0.006 & 0.064 & V &-  &  10.459$\pm$0.009 & 0.068 &  V    &-                             &2.06&          V     \\
IC1396A-50 &21:37:07.18 +57:31:27.8 &I      &           &I & 13.585$\pm$0.016 & 0.04  & V &- & 12.718$\pm$0.017 & 0.027 & V   & -                                 &0.21&          V     \\
IC1396A-51\tablenotemark{e} &21:37:07.71 +57:32:11.0 &II  &           &II & 12.923$\pm$0.012 & 0.042 & V &- & 12.666$\pm$0.016 & 0.052 & V   &-                   &0.23&          N     \\
IC1396A-53 &21:37:09.36 +57:29:48.3 &II &           &II & 10.973$\pm$0.006 & 0.019 & V &- & 10.515$\pm$0.009 & 0.028 & V   & -                                    &0.74&          V     \\
IC1396A-54 &21:37:09.44 +57:30:36.7 &II     &           &II & 12.125$\pm$0.009 & 0.035 & V &- & 11.817$\pm$0.009 & 0.047 & V   & -                                &0.72&          V     \\
IC1396A-55 &21:37:10.13 +57:31:26.6 &II     &           &II & 13.252$\pm$0.015 & 0.029 & N &- & 12.918$\pm$0.019 & 0.029 & N   & -                                &0.24&          N     \\
IC1396A-56 &21:37:10.31 +57:30:18.9 &II     &           &II & 11.545$\pm$0.009 & 0.011 & N &- & 11.417$\pm$0.009 & 0.017 & N   & -                                &0.17&          N     \\
IC1396A-57 &21:37:10.54 +57:31:12.4 &II     &           &II & 11.130$\pm$0.006 & 0.058 & V &- & 10.724$\pm$0.009 & 0.063 & V   & -                                &2.19&          V     \\
IC1396A-58 &21:36:56.51 +57:31:51.6 &--     &$\kappa$,II    &II & 12.323$\pm$0.009 & 0.016 & N &- & 11.626$\pm$0.009 & 0.015 & N   &-                             &0.05&          N     \\
IC1396A-59 &21:37:03.04 +57:30:48.7 &--     &$\mu$,I    &I & 13.564$\pm$0.016 & 0.032 & N &- & 12.218$\pm$0.013 & 0.036 & N   &-                                  &0.48&          N     \\
IC1396A-60 &21:36:47.18 +57:29:52.6 &--     &           &II & 10.406$\pm$0.006 & 0.062 & V &- &  9.869$\pm$0.009 & 0.04  & V   &-                                 &1.83&          V     \\
IC1396A-61 &21:36:47.63 +57:29:54.1 &--     &           &II & 10.416$\pm$0.006 & 0.030 & V &- &  9.960$\pm$0.009 & 0.033 & V   &-                                 &1.30&          V     \\
IC1396A-62 &21:37:14.51 +57:28:40.9 &--     &           &II & 11.833$\pm$0.009 & 0.061 & V &- & -- & - & -   &-                                                   &1.96&          V     \\
IC1396A-63 &21:37:17.42 +57:29:27.5 &--     &           &II & 11.835$\pm$0.009 & 0.038 & V &- & -- & -- & -   &-                                                  &1.01&          V     \\
IC1396A-64 &21:37:17.37 +57:29:20.8 &--     &           &II & 12.886$\pm$0.012 & 0.019 & N &- & -- & -- & -   &-                                                  &0.10&          N     \\
IC1396A-65 &21:36:42.49 +57:25:23.3 &--     &           &II & 13.176$\pm$0.015 & 0.018 & N &- & 12.943$\pm$0.019 & 0.036 & N   &-                                 &0.16&          N     \\
IC1396A-66 &21:37:05.87 +57:32:12.5 &--     &           &II & 13.362$\pm$0.015 & 0.044 & V &9 & 13.074$\pm$0.021 & 0.033 & N   &-                                 &0.25&          V     \\
IC1396A-67 &21:35:58.52 +57:29:15.1 &--     &           &II & 13.546$\pm$0.016 & 0.036 & N &- & 13.098$\pm$0.021 & 0.027 & N   &-                                 &0.15&          N     \\
IC1396A-68 &21:36:56.26 +57:29:52.3 &--     &           &II & 13.551$\pm$0.016 & 0.023 & N &- & 12.974$\pm$0.02  & 0.024 & N   &-                                 &0.09&          N     \\
IC1396A-69 &21:36:38.03 +57:26:57.9 &--     &           &II & 13.578$\pm$0.016 & 0.048 & V &- & 13.172$\pm$0.022 & 0.077 & V   &-                                 &0.90&          V     \\
IC1396A-70 &21:36:12.60 +57:31:26.3 &--     &           &II & 13.997$\pm$0.021 & 0.126 & V &- & 13.473$\pm$0.026 & 0.095 & V   &-                                 &1.08&          V     \\
IC1396A-71 &21:36:40.34 +57:25:45.7 &--     &           &II& 14.038$\pm$0.022 & 0.043 & N &-  &  13.617$\pm$0.028 & 0.042 &  N   &-                               &0.34&          N     \\
IC1396A-72 &21:36:18.97 +57:29:05.1 &--     &           &I & 14.186$\pm$0.024 & 0.051 & V &-  &   -- & -- & -    &-                                               &0.13&          N     \\
IC1396A-73 &21:37:11.78 +57:30:34.9 &--     &           &II & 14.269$\pm$0.026 & 0.051 & N &-  &  13.823$\pm$0.031 & 0.051 &  N   &-                              &0.26&          N     \\
IC1396A-74 &21:36:36.35 +57:32:09.3 &--     &           &I & 15.491$\pm$0.059 & 0.053 & N &-  &  14.663$\pm$0.050 & 0.040 &  N   &-                               &-0.05&          V     \\
\enddata
\tablenotetext{a}{Objects are named following the order in Table~6 of \citet{Sicilia06} (Note that not all their members are included), objects -58 and -59 are YSOSs from \citet{Reach04} not in common with \citet{Sicilia06}, and objects from -60 to -74 are our new candidate members from Table~\ref{tab:newphot}.}
\tablenotetext{b}{YSO Class as in \citet{Sicilia06}.}
\tablenotetext{c}{ID and YSO Class as in \citet{Reach04}.}
\tablenotetext{d}{YSO Class in this paper. Members in \citet{Sicilia06} that do not show IR excess in our data have been classified as Class~III.}
\tablenotetext{e}{Uncertain cluster members following \citet{Sicilia06}.}
\tablenotetext{f}{Classification in disagreement with \citet{Sicilia06}.}
\end{deluxetable}


\clearpage
\begin{deluxetable}{lccccc}
\tabletypesize{\normalsize}
\tablecolumns{6}
\tablewidth{0pt}
\tablecaption{Time series in the four IRAC bands for the variable YSOs of our sample. (Full table available in the electronic version of this article.) \label{tab:timeseries}}
\tablehead{ 
\colhead{Object} &\colhead{MJD (days)} &\colhead{[3.6] e[3.6]} &\colhead{[4.5] e[4.5]} &\colhead{[5.8] e[5.8]} &\colhead{[8.0] e[8.0]}}
\startdata
IC1396A-1  &  54488.6878678 &  13.233 0.018 &  12.596 0.013 &  11.833 0.023 &  10.448 0.042 \\
IC1396A-1  &  54489.2077075 &  13.149 0.014 &  12.545 0.012 &  11.741 0.028 &  10.365 0.037 \\
IC1396A-1  &  54489.5421136 &  13.178 0.019 &  12.529 0.011 &  11.730 0.020 &  10.415 0.038 \\
IC1396A-1  &  54490.0224986 &  13.229 0.019 &  12.583 0.012 &  11.842 0.026 &  10.405 0.037 \\
IC1396A-1  &  54490.3616676 &  13.212 0.020 &  12.572 0.012 &  11.760 0.019 &  10.360 0.035 \\
IC1396A-1  &  54490.7798051 &  13.256 0.021 &  12.585 0.011 &  11.800 0.024 &  10.340 0.035 \\
IC1396A-1  &  54491.2620497 &  13.293 0.021 &  12.682 0.014 &  11.867 0.024 &  10.501 0.036 \\
IC1396A-1  &  54491.6974542 &  13.384 0.023 &  12.774 0.014 &  12.000 0.030 &  10.531 0.040 \\
IC1396A-1  &  54492.3575716 &  13.428 0.027 &  12.836 0.014 &  12.065 0.030 &  10.604 0.036 \\
IC1396A-1  &  54492.8026341 &  13.424 0.028 &  12.783 0.013 &  12.037 0.030 &  10.633 0.039 \\
\enddata

\end{deluxetable}

\clearpage

\begin{deluxetable}{lcccc}
\tabletypesize{\scriptsize}
\tablecaption{Peak-to-peak amplitudes for the 18 IC 1396A YSOs which present periodic-like, colorless variations. Long term data.\label{tab:amp}}
\tablewidth{0pt}
\tablehead{
\colhead{}      & \multicolumn{4}{c}{Peak-to-peak Amplitudes (mag)} \\
\colhead{Object}          & \colhead{3.6~$\mu$m}            & \colhead{4.5~$\mu$m}   & \colhead{5.8~$\mu$m}     & \colhead{8.0~$\mu$m} 
}
\startdata
IC~1396A-4   & 0.15 & 0.11 & 0.09 & 0.07 \\
IC~1396A-6\tablenotemark{a}  & 0.09 & 0.07 & 0.15 & --   \\
IC~1396A-7   & 0.19 & 0.19 & 0.17 & 0.15 \\
IC~1396A-13\tablenotemark{a} & 0.17 & 0.17 & 0.23 & --   \\
IC~1396A-14  & 0.09 & 0.11 & 0.14 & 0.15 \\
IC~1396A-15\tablenotemark{b} & 0.12 & 0.12 & 0.49 & 0.53 \\
IC~1396A-16\tablenotemark{b} & 0.09 & 0.09 & 0.2  & 0.15 \\ 
IC~1396A-17  & 0.10 & 0.10 & 0.10 & 0.11 \\	
IC~1396A-22  & 0.20 & 0.18 & 0.21 & 0.23 \\
IC~1396A-24  & 0.11 & 0.12 & 0.11 & 0.11 \\
IC~1396A-28\tablenotemark{b} & 0.15 & 0.12 & 0.16 & 0.23 \\
IC~1396A-29  & 0.06 & 0.06 & 0.04 & 0.05 \\
IC~1396A-34  & 0.11 & 0.10 & 0.12 & 0.08 \\
IC~1396A-37  & 0.14 & 0.13 & 0.12 & 0.17 \\
IC~1396A-39  & 0.17 & 0.16 & 0.19 & 0.22 \\
IC~1396A-40  & 0.07 & 0.07 & 0.06 & 0.06 \\
IC~1396A-41\tablenotemark{b} & 0.06 & 0.07 & 0.23 & 0.51 \\
IC~1396A-43  & 0.16 & 0.16 & 0.17 & 0.18 \\
IC~1396A-45  & 0.07 & 0.07 & 0.07 & 0.06 \\
IC~1396A-47  & 0.25 & 0.21 & 0.24 & 0.25 \\
IC~1396A-60  & 0.13 & 0.11 & 0.12 & 0.07 \\
IC~1396A-61  & 0.08 & 0.09 & 0.11 & 0.09 \\
IC~1396A-66\tablenotemark{b} & 0.08 & 0.11 & 0.25 & 0.3  \\
\enddata
\tablenotetext{a}{This object is too faint to derive its amplitude at 8.0$\mu$m.}
\tablenotetext{b}{The light curves of this object are much noisier at 5.8 and 8.0$\mu$m than at shorter wavelenths}
\end{deluxetable}


\clearpage

\begin{deluxetable}{lcccccccccc}
\tabletypesize{\scriptsize}
\tablecaption{Short-time variability: Main results from the staring data.\label{tab:shortvar}}
\tablewidth{0pt}
\tablehead{
\colhead{}      & \multicolumn{4}{c}{4.5~$\mu$m} &  \multicolumn{4}{c}{8~$\mu$m}& \colhead{} & \colhead{}\\
\colhead{Object} & \multicolumn{4}{c}{---------------------------------------------------} &  \multicolumn{4}{c}{---------------------------------------------------}& \colhead{J index} & \colhead{Var}\\
\colhead{}          & \colhead{mag}            & \colhead{RMS}   & \colhead{$\chi^2$}     & \colhead{$T_{var}$ (hr)} & \colhead{mag}            & \colhead{RMS}   & \colhead{$\chi^2$}     & \colhead{$T_{var}$ (hr)}& \colhead{}& \colhead{Fin}
}
\startdata
IC~1396-16 & 10.910$\pm$0.006 & 0.009 & N &- &10.246$\pm$0.181 & 0.256 & N&-& 0.03 & N\\
IC~1396-17 & 10.600$\pm$0.003 & 0.006 & V &- & 9.167$\pm$0.011 & 0.014 & N&-& 0.07 & N\\
IC~1396-18 & 10.970$\pm$0.004 & 0.006 & V &- &10.018$\pm$0.042 & 0.057 & N&-& -0.11& N\\
IC~1396-20 & 12.368$\pm$0.007 & 0.010 & N &- &11.323$\pm$0.068 & 0.093 & N&-& 0.02 & N\\
IC~1396-21 & 12.143$\pm$0.007 & 0.010 & N &- &10.180$\pm$0.029 & 0.040 & N&-& 0.09 & N\\
IC~1396-22 & 10.585$\pm$0.003 & 0.033 & V &- & 9.742$\pm$0.031 & 0.046 & N&-& 1.69 & Y\\
IC~1396-23 & 12.392$\pm$0.009 & 0.012 & N &- &10.284$\pm$0.039 & 0.050 & N&-& 0.05 & N\\
IC~1396-24 &  9.008$\pm$0.002 & 0.005 & V &- & 8.000$\pm$0.004 & 0.005 & N&-& 0.37 & Y\\
IC~1396-25 &  8.739$\pm$0.002 & 0.003 & N &- & 6.614$\pm$0.002 & 0.006 & V&-& 0.56 & N\\
IC~1396-28 & 11.910$\pm$0.015 & 0.018 & N &- & 8.558$\pm$0.037 & 0.053 & N&-& -0.01& N\\
IC~1396-29 &  9.009$\pm$0.002 & 0.003 & N &- & 7.941$\pm$0.005 & 0.007 & N&-& 0.09& N\\
IC~1396-31 & 12.245$\pm$0.006 & 0.009 & N &- &11.187$\pm$0.043 & 0.060 & N&-& 0.01& N\\
IC~1396-32 &  9.870$\pm$0.002 & 0.003 & N &- & 9.690$\pm$0.023 & 0.032 & N&-& 0.01& N\\
IC~1396-33 & 13.251$\pm$0.018 & 0.024 & N &- &10.602$\pm$0.126 & 0.173 & N&-& 0.11& N\\
IC~1396-34 & 11.266$\pm$0.005 & 0.012 & V &- & 9.749$\pm$0.009 & 0.013 & N&-& 0.42& N\\
IC~1396-35 &  9.263$\pm$0.002 & 0.009 & V &- & 8.027$\pm$0.007 & 0.014 & V&-& 1.70& Y\\
IC~1396-36 & 11.230$\pm$0.004 & 0.005 & N &- &10.811$\pm$0.080 & 0.108 & N&-& -0.07& N\\
IC~1396-37 & 11.027$\pm$0.004 & 0.006 & V &- & 9.931$\pm$0.020 & 0.025 & N&-& 0.02& N\\
IC~1396-39 & 10.979$\pm$0.004 & 0.018 & V &- & 8.898$\pm$0.033 & 0.046 & N&-& 0.56& Y\\
IC~1396-40 &  9.917$\pm$0.002 & 0.003 & N &- & 8.719$\pm$0.005 & 0.007 & N&-& 0.08& N\\
IC~1396-41 & 12.193$\pm$0.008 & 0.011 & N &- &10.452$\pm$0.093 & 0.141 & N&-& 0.084& N\\
IC~1396-43 & 11.230$\pm$0.005 & 0.008 & V &- & 9.564$\pm$0.022 & 0.035 & N&-& 0.34& Y\\
IC~1396-44 & 13.116$\pm$0.017 & 0.024 & N &- & 9.001$\pm$0.017 & 0.021 & -&-& 0.03& N\\
IC~1396-46 & 11.912$\pm$0.008 & 0.010 & N &- &--               & --    & -&-& 0.22& N\\
IC~1396-47 &  9.722$\pm$0.003 & 0.011 & V &3.4&8.115$\pm$0.007&0.014 & V&3.4& 1.45& Y\\
IC~1396-48 & 13.581$\pm$0.060 & 0.080 & N &- & 8.846$\pm$0.044 & 0.059 & N&-& -0.04& N\\
IC~1396-50 & 12.689$\pm$0.008 & 0.010 & N &- &10.128$\pm$0.013 & 0.017 & N&-& 0.07& N\\
IC~1396-51\tablenotemark{a}&--& --    & - &- &12.278$\pm$0.047 & 0.063 & N&-& -0.14& N\\
IC~1396-54 & 11.682$\pm$0.004 & 0.008 & V &- &--               & --    & -&-& 0.93& N\\
IC~1396-55 & 12.918$\pm$0.008 & 0.011 & N &- &11.898$\pm$0.036 & 0.049 & N&-& 0.02& N\\
IC~1396-57 & 10.628$\pm$0.003 & 0.005 & V &- & 9.458$\pm$0.004 & 0.008 & V&-& 0.56& Y\\
IC~1396-58 & 11.586$\pm$0.004 & 0.006 & N &- &10.088$\pm$0.027 & 0.037 & N&-& 0.06& N\\
IC~1396-59 & 12.197$\pm$0.006 & 0.007 & N &- &10.790$\pm$0.036 & 0.049 & N&-& -0.03& N\\
IC~1396-60 &  9.867$\pm$0.003 & 0.005 & V &- &--               & --    & -&-& 0.89& N\\
IC~1396-61 &  9.923$\pm$0.003 & 0.007 & V &- & 8.969$\pm$0.009 & 0.015 & V&-& 0.82& Y\\
IC~1396-66 & 13.038$\pm$0.010 & 0.013 & N &- &11.714$\pm$0.041 & 0.058 & N&-& -0.14& N\\
IC~1396-68 & 13.023$\pm$0.013 & 0.017 & N &- &13.107$\pm$0.474 & 0.679 & N&-& -0.09& N\\
IC~1396-74 & 14.663$\pm$0.038 & 0.050 & N &- &11.648$\pm$0.063 & 0.080 & N&-& -0.01& N\\
\enddata
\tablenotetext{a}{The object is out of the FOV in Ch.~2.}
\end{deluxetable}


\clearpage
\begin{deluxetable}{lcccccccc}
\tabletypesize{\scriptsize}
\rotate
\tablecolumns{16}
\tablewidth{0pt}
\tablecaption{Photometry for the variable objects found in the field.\label{tab:fieldvariables}}
\tablehead{ 
\colhead{ID\tablenotemark{a}} &\colhead{RA(J2000), DEC(J2000)} &\colhead{J\tablenotemark{b} eJ} &\colhead{H\tablenotemark{b} eH} &\colhead{$K_s$\tablenotemark{b} e$K_s$}&\colhead{[3.6] e[3.6]} &\colhead{[4.5] e[4.5]} &\colhead{[5.8] e[5.8]} &\colhead{[8.0] e[8.0]}}
\startdata
  347  &21:35:51.08 +57:28:12.4 &15.381 0.033  &14.881 0.042  &14.611 0.038  &14.272    0.029   &14.322    0.027   &13.268    0.135   &11.331    0.095\\  
  751  &21:36:15.20 +57:25:27.8 &15.442 0.047  &14.659 0.052  &14.418 0.038  &13.787    0.01    &13.556    0.012   &13.585    0.063   &12.365    0.06 \\  
  823  &21:36:07.44 +57:26:43.4 &--     --     &--     --     &--     --     &12.869    0.035   &11.032    0.025   &9.635     0.019   &8.576     0.055\\  
  1230 &21:35:58.07 +57:28:50.4 &15.596 0.088  &15.059 0.125  &14.836 0.091  &13.899    0.034   &14.067    0.055   &13.809    0.18    &--        --   \\  
  1256 &21:35:53.10 +57:29:37.1 &15.208 0.031  &14.709 0.047  &14.499 0.036  &14.424    0.025   &14.407    0.027   &15.228    0.417   &14.529    0.908\\  
  1469 &21:35:55.39 +57:29:42.8 &15.654 0.034  &15.059 0.042  &14.857 0.049  &14.688    0.028   &14.673    0.032   &14.603    0.207   &--        --   \\  
  1685 &21:35:55.64 +57:30:03.3 &15.314 0.03   &14.767 0.041  &14.453 0.038  &14.136    0.014   &14.151    0.019   &14.034    0.087   &--        --   \\  
  2519 &21:36:01.65 +57:30:50.0 &15.598 0.041  &14.957 0.052  &14.701 0.041  &14.146    0.017   &14.02     0.018   &14.082    0.137   &13.755    0.269\\  
  2712 &21:36:10.97 +57:29:50.7 &15.649 0.034  &14.601 0.037  &14.256 0.027  &14.11     0.019   &14.067    0.022   &14.204    0.319   &13.246    0.835\\  
  2812 &21:36:40.34 +57:25:45.6 &16.292 0.049  &15.439 0.063  &14.862 0.04   &14.102    0.015   &13.688    0.016   &13.187    0.042   &12.506    0.047\\  
  3115 &21:36:13.35 +57:30:16.2 &15.697 0.039  &15.111 0.052  &14.854 0.05   &14.716    0.039   &14.635    0.035   &14.753    0.342   &--        --   \\  
  3154 &21:36:17.04 +57:29:48.2 &15.674 0.034  &14.63  0.041  &14.198 0.029  &13.789    0.048   &13.741    0.053   &14.619    1.782   &  --       --  \\  
  3367 &21:36:33.20 +57:27:51.9 &14.985 0.028  &14.413 0.04   &14.208 0.031  &14.001    0.022   &13.874    0.018   &14.172    0.088   &--        --   \\  
  3402 &21:36:16.14 +57:30:26.9 &14.598 0.025  &14.209 0.036  &14.006 0.026  &13.909    0.017   &13.797    0.014   &13.635    0.119   &12.549    0.248\\  
  3444 &21:36:17.94 +57:30:16.3 &15.531\tablenotemark{c} --     &15.35  0.07   &15.048 0.057  &14.474    0.034   &14.419    0.033   &--        --      &--        --   \\  
  3526 &21:36:45.86 +57:26:22.9 &15.174 0.03   &14.494 0.043  &14.339 0.033  &14.264    0.012   &14.326    0.024   &14.171    0.091   &13.071    0.073\\  
  3795 &21:36:33.02 +57:28:49.4 &16.154 0.048  &14.919 0.046  &14.127 0.032  &13.738    0.04    &13.344    0.037   &12.895    0.354   &12.874    1.697\\  
  4459 &21:36:36.94 +57:29:28.6 &15.715 0.03   &15.014 0.04   &14.599 0.041  &14.267    0.076   &14.173    0.11    &12.556    0.322   &--        --   \\  
  4535 &21:36:37.58 +57:29:31.7 &15.468 0.03   &14.983 0.038  &14.816 0.045  &14.444    0.078   &14.431    0.105   &12.978    0.176   &11.433    0.221\\  
  4780 &21:36:54.72 +57:27:26.7 &--     --     &--     --     &--     --     &16.999    0.109   &16.836    0.255   &--        --      &--        --   \\  
  5053 &21:36:12.99 +57:34:05.4 &15.167 0.03   &14.053 0.032  &13.752 0.024  &13.474    0.022   &13.498    0.018   &13.355    0.161   &13.58     0.606\\  
  5120 &21:36:38.61 +57:30:27.0 &17.113 0.103  &15.528 0.051  &14.937 0.051  &14.873    0.134   &14.708    0.145   &--        --      &--        --   \\  
  5272 &21:36:40.49 +57:30:25.8 &15.99  0.052  &14.524 0.053  &13.664\tablenotemark{c} --     &13.32     0.043   &13.225    0.061   &12.12     0.116   &10.638    0.172\\  
  5290 &21:36:28.43 +57:32:13.5 &14.431 0.023  &13.482 0.03   &13.144 0.019  &12.901    0.013   &12.824    0.012   &13.889    0.321   &--        --   \\  
  5546 &21:36:47.62 +57:29:54.0 &13.497 0.024  &12.356 0.038  &11.593 0.019  &10.477    0.004   &10.017    0.004   &9.672     0.012   &9.082     0.036\\  
  5580 &21:36:25.98 +57:33:10.2 &14.883 0.029  &14.265 0.043  &14.011 0.029  &13.702    0.037   &13.638    0.025   &14.196    0.629   &  --       --  \\  
  5610 &21:36:49.03 +57:29:49.2 &16.904 0.079  &15.51  0.059  &14.648 0.043  &14.157    0.043   &14.126    0.072   &15.489    2.214   &--        --   \\  
  5647 &21:36:44.11 +57:30:38.3 &16.214 0.056  &15.237 0.06   &14.754 0.045  &14.431    0.085   &14.431    0.075   &14.023    0.386   &--        --   \\  
  5676 &21:36:44.71 +57:30:37.4 &16.445 0.053  &15.464 0.063  &14.851 0.043  &14.388    0.097   &14.501    0.094   &--        --      &11.127    0.166\\  
  5680 &21:36:16.09 +57:34:48.6 &11.97  0.022  &11.185 0.028  &10.931 0.016  &10.791    0.004   &10.746    0.004   &10.77     0.01    &10.684    0.01 \\  
  5998 &21:36:45.86 +57:31:03.5 &--     --     &--     --     &--     --     &14.223    0.106   &14.578    0.146   &--        --      &10.166    0.294\\  
  6200 &21:36:53.17 +57:30:18.7 &17.607\tablenotemark{c} --     &16.315 0.123  &15.177 0.067  &14.426    0.073   &14.118    0.056   &14.5      1.015   &  --       --  \\  
  6450 &21:36:36.50 +57:33:14.3 &16.876 0.079  &16.196 0.125  &15.781 0.092  &14.815    0.116   &14.933    0.098   &--        --      &--        --   \\  
  6710 &21:36:34.86 +57:33:57.1 &15.135 0.026  &14.478 0.031  &14.251 0.031  &14.171    0.016   &14.175    0.021   &14.063    0.087   &--        --   \\  
  6975 &21:36:25.61 +57:35:46.2 &--     --     &--     --     &--     --     &14.152    0.058   &14.068    0.051   &14.586    0.365   &--        --   \\  
  7003 &21:36:52.64 +57:31:50.4 &15.005 0.029  &13.703 0.032  &13.156 0.019  &12.673    0.024   &12.623    0.016   &11.816    0.115   &10.26     0.157\\  
  7159 &21:36:54.58 +57:31:50.2 &17.258 0.108  &15.771 0.074  &15.035 0.05   &14.067    0.076   &13.879    0.065   &13.74     0.721   &12.97     1.429\\  
  7377 &21:37:10.55 +57:29:52.9 &15.107 0.028  &14.3   0.033  &14.134 0.027  &13.85     0.013   &13.842    0.027   &13.849    0.11    &14.044    0.348\\  
  7491 &21:36:51.54 +57:32:53.4 &16.231 0.046  &15.285 0.055  &14.846 0.047  &13.917    0.092   &13.936    0.072   &13.172    0.754   &11.122    0.534\\  
  7493 &21:36:48.84 +57:33:17.4 &17.542 0.122  &16.693 0.154  &16.584 0.21   &14.666    0.086   &14.931    0.087   &--        --      &--        --   \\  
  7532 &21:36:54.66 +57:32:29.4 &17.35\tablenotemark{c}  --     &16.177 0.105  &15.27  0.08   &13.858    0.075   &13.85     0.052   &11.82     0.127   &9.94      0.124\\  
  7949 &21:37:00.27 +57:32:23.8 &--     --     &--     --     &--     --     &14.256    0.256   &14.874    0.341   &11.454    0.127   &9.457     0.129\\  
  8151 &21:36:45.97 +57:34:55.1 &15.102 0.047  &14.434 0.059  &14.254 0.036  &13.811    0.032   &13.949    0.061   &14.221    0.22    &--        --   \\  
  8183 &21:36:59.85 +57:32:56.2 &15.636 0.037  &15.23  0.049  &15.003 0.053  &14.843    0.048   &14.686    0.049   &14.55     0.323   &--        --   \\  
  8507 &21:37:06.49 +57:32:32.1 &13.33  0.02   &12.335 0.028  &11.904 0.016  &10.862    0.0030  &10.539    0.0030  &10.201    0.0070  &9.527     0.009\\ 
  8689 &21:37:09.39 +57:32:25.4 &15.82  0.116  &15.205 0.181  &15.016 0.07   &13.793    0.052   &13.394    0.091   &13.299    0.135   &--        --   \\  
\enddata
\tablenotetext{a}{Internal ID for each object.}
\tablenotetext{b}{Photometry for the J, H and $K_s$ bandpasses has been taken from the 2MASS database \citep{Cutri03}.}
\tablenotetext{c}{Upper limmits.}
\end{deluxetable}

\clearpage

\begin{deluxetable}{lcc}
\tabletypesize{\small}
\tablecaption{Statistics for the long term variability.\label{tab:statistics}}
\tablewidth{0pt}
\tablehead{
\colhead{Sample}          & \colhead{Number of objects} & \colhead{\% Total\tablenotemark{a}}
}
\startdata
Class I in the sample & 20 & 29\\
Class II in the sample & 45 & 65\\
Variable objects & 41 & 59\\
Class I with amplitudes $>$ 0.05mag & 15 & 22\\
Class II with amplitudes $>$ 0.05mag & 26 & 38\\
Class I with amplitudes $>$ 0.1mag & 11 & 16\\
Class II with amplitudes $>$ 0.1mag & 17 & 25\\
Periodic-like Class I & 9 & 13\\
Periodic-like Class II & 14 & 20\\
Color variables & 3 & 4\\
\enddata
\tablenotetext{a}{Percentage to the total sample (69 objects).}
\end{deluxetable}


\clearpage

\clearpage

\clearpage

\clearpage

\clearpage

\clearpage


\clearpage

\clearpage

\end{document}